# Title: Electronic-grade epitaxial (111) KTaO$_3$ heterostructures

**Authors:** Jieun Kim[1,†], Muqing Yu[2,3,†], Jung-Woo Lee[1,†], Shun-Li Shang[4], Gi-Yeop Kim[5], Pratap Pal[1], Jinsol Seo[6,7], Neil Campbell[8], Kitae Eom[1], Ranjani Ramachandran[2,3], Mark S. Rzchowski[8], Sang Ho Oh[6,7], Si-Young Choi[5], Zi-Kui Liu[4], Jeremy Levy[2,3], Chang-Beom Eom[1,*]

**Affiliations:**

[1]Department of Materials Science and Engineering, University of Wisconsin-Madison, Madison, WI 53706, USA.

[2]Department of Physics and Astronomy, University of Pittsburgh, Pittsburgh, PA 15260, USA.

[3]Pittsburgh Quantum Institute, Pittsburgh, PA, 15260, USA.

[4]Department of Materials Science and Engineering, The Pennsylvania State University, University Park, Pennsylvania 16802, USA.

[5]Department of Materials Science and Engineering, Pohang University of Science and Technology77 Cheongam-Ro, Pohang 37673, Republic of Korea.

[6]Department of Energy Science, Sungkyunkwan University, Suwon 16419, Korea.

[7]Department of Energy Engineering, KENTECH Institute for Energy Materials and Devices, Korea Institute of Energy Technology (KENTECH), Naju 58330, Republic of Korea.

[8]Department of Physics, University of Wisconsin-Madison, Madison, WI, 53706, USA.

†These authors contributed equally to this work.

*Corresponding author. Email: ceom@wisc.edu

**Abstract:** KTaO$_3$ has recently attracted attention as a model system to study the interplay of quantum paraelectricity, spin-orbit coupling, and superconductivity. However, the high and low vapor pressures of potassium and tantalum present processing challenges to creating interfaces clean enough to reveal the intrinsic quantum properties. Here, we report superconducting heterostructures based on electronic-grade epitaxial (111) KTaO$_3$ thin films. Electrical and structural characterizations reveal that two-dimensional electron gas at the heterointerface between amorphous LaAlO$_3$ and KTaO$_3$ thin film exhibits significantly higher electron mobility, superconducting transition temperature and critical current density than those in bulk single crystal KTaO$_3$-based heterostructures owing to cleaner interface in KTaO$_3$ thin films. Our hybrid approach may enable epitaxial growth of other alkali metal-based oxides that lie beyond the capabilities of conventional methods.

**One-Sentence Summary:** Hybrid growth design yields electronic-grade epitaxial quantum heterostructures containing volatile and refractory elements.

**Main Text:** Quantum materials made of transition metal oxides display numerous interesting physical properties such as magnetism, ferroelectricity, interfacial conductivity, and superconductivity (*1–5*). In their epitaxial thin film forms, the strong *d* electron correlations that determine many of their physical properties can be easily manipulated through coupling to strain, dimensionality, and chemical pressures (*6–8*). Traditionally, pulsed-laser deposition (PLD) and molecular-beam epitaxy (MBE) have been workhorse synthesis tools in the quest to produce a model thin film version of novel quantum materials exhibiting exotic quantum phenomena (*9–11*).

In PLD, a target having the desired composition as the growing film is ablated by a pulsed-laser beam, creating a plume consisting of a stoichiometric mixture of high-energetic ions. In contrast, each element is thermally evaporated from separate effusion cells, creating a low-energetic molecular beam of evaporated atoms in MBE. Although both are excellent techniques with wide-ranging applications, numerous interesting materials still lie outside their capabilities (*12–14*). Examples include perovskite oxides with a general formula of $A^{1+}B^{5+}O_3$, where the A-site is usually occupied by alkali metals (*e.g.*, lithium, potassium, etc.) while the B-site is occupied by refractory metals (*e.g.*, niobium, tantalum, etc.), such as potassium tantalate ($KTaO_3$). In these compounds, an extremely large vapor pressure mismatch of their constituents (fig. S1) presents considerable processing challenges for growing atomically precise and stoichiometric thin films. To alleviate these issues, approaches such as potassium-enriched ablation targets (*15*) in PLD and suboxide tantalum sources (*16*) in MBE have been put forth. Notwithstanding such advances, $A^{1+}B^{5+}O_3$ perovskites and other alkali-metal containing compounds (*e.g.*, $NaNbO_3$, $KV_3Sb_5$) remain as one of the least explored materials systems in thin film forms.

In this work, we describe an alternative approach, which synergistically combines the advantages of PLD and MBE, namely hybrid PLD. We synthesize electronic-grade epitaxial thin films of (111)-oriented $KTaO_3$ and analyze superconducting two-dimensional electron gases (2DEGs) at its interface with $LaAlO_3$. $KTaO_3$ on its own hosts many interesting physical properties such as quantum paraelectricity and large spin-orbit coupling (*17–19*). Bulk single crystal $KTaO_3$ (111)-based heterointerfaces including $EuO/KTaO_3$ (111) (*4*), $LaTiO_3/KTaO_3$ (111) (*20*), $La_{2/3}Sr_{1/3}MnO_3/KTaO_3$ (111) (*21*), $AlO_x/KTaO_3$ (111) (*22*), and $LaAlO_3/KTaO_3$ (111) (*5*) have displayed superconductivity with a high superconducting transition temperature ($T_c$) ≈1-2 K, with its properties inherently correlated with the crystallographic orientation (*23*). Reducing the point defect concentrations below that of bulk single crystals is a key challenge for studying the origin of this superconductivity, and related quantum phenomena (*19*) such as ferroelectric quantum criticality (*24–26*), topological superconductivity (*27*) and superconducting spintronics (*28*). Using a combination of thermodynamic analysis (*29*) and synthesis design (*30*), we are able to grow electronic-grade homoepitaxial $KTaO_3$ thin films on buffered single crystal $KTaO_3$ (111) substrates, followed by an *in situ* $LaAlO_3$ growth by conventional PLD.

**Adsorption-controlled hybrid synthesis approach**

To predict the growth parameters of a stoichiometric $KTaO_3$ phase, we built a thermodynamic database for the K-Ta-O system (*31*) and predicted relevant phase diagrams (fig. S2-S4). Fig. 1A shows the calculated stability phase diagram of K-Ta-O near the stoichiometric $KTaO_3$ as a function of K partial pressure and temperature (see fig. S3 for a complete phase diagram of Fig. 1A); here, the oxygen ($O_2$) partial pressure is fixed at $10^{-6}$ Torr based on the potential phase diagram as a function of K and $O_2$ partial pressures (fig. S4). In our synthesis we used commercially available potassium oxide ($K_2O$) as K source because elemental K is extremely unstable in both the ambient and high temperature/vacuum conditions (*31*). We estimate the source temperatures for achieving stoichiometric $KTaO_3$ synthesis by calculating partial pressures of all gas species (Fig. 1B) at source temperatures from 500-1000 K. We identify three major gas species as $O_2$, $K_2O_2$, and K with the calculated equilibrium partial pressures ≈$1.7\times10^{-3}$, $9.2\times10^{-7}$, and $3.9\times10^{-7}$ Torr, respectively, at a source temperature of 750 K. These values set the upper limits for the partial pressures of gas species.

In our experiments, the vacuum chamber is constantly evacuated, maintaining dynamic equilibrium, so that the total dynamic pressure from the combined gas phases at a source temperature of 750 K comes to ≈$10^{-6}$ Torr (*31*). The chamber pressure was kept well below $10^{-3}$ Torr to maintain the molecular flow state of gas species and sufficient overpressure of potassium near the surface of substrate. These analyses indicated that the practical growth window of KTaO$_3$ should be in the range of $10^{-7}$-$10^{-9}$ Torr K partial pressure and 950-1000 K substrate temperatures (red box, Fig. 1A). Fig. 1C schematically depicts the hybrid PLD experimental setup (*31*). As with MBE, K is supplied by thermal evaporation of a K$_2$O effusion cell directed at the substrate. In contrast, Ta is supplied by ablating a ceramic target of tantalum pentoxide (Ta$_2$O$_5$) with a pulsed excimer laser as in PLD. Fig. 1D schematically illustrates the adsorption-controlled growth of KTaO$_3$. In adsorption-controlled growth (*32*), the volatile species (K in this case) is provided with sufficiently large overpressure to avoid for K deficiency while excess K readily evaporates from the K-terminated surface. Although KTaO$_3$ melts incongruently (*33*) and hence perfect stoichiometry may still be challenging, the proximity to thermodynamic equilibrium afforded by the adsorption-controlled growth minimizes unintentional defects and hence facilitates the synthesis of electronic-grade KTaO$_3$ thin films.

**Epitaxial thin film synthesis**

We grew ≈8-10 nm-thick epitaxial KTaO$_3$ thin films on single crystal substrates of SrTiO$_3$ (001), SrTiO$_3$ (111), KTaO$_3$ (111), and on KTaO$_3$ (111) with an ≈1 nm-thick SmScO$_3$ template layer (fig. S5-7). In the main text, we only discuss the films grown homoepitaxially on KTaO$_3$ (111) for brevity. Homoepitaxial KTaO$_3$ thin films grown directly on bare KTaO$_3$ (111) substrate at substrate temperature of 973 K show atomically smooth surfaces (fig. S7B), but X-ray diffraction indicate a slight off-stoichiometry of the KTaO$_3$ thin film (fig. S7H). An increased substrate temperature of 1023 K enhances stoichiometry, but results in surface roughening generating the more thermodynamically stable (001) and (011) facets (fig. S7C) (*34*, *35*). To address these issues, we grew a thin SmScO$_3$ template on KTaO$_3$ (111) substrates and grew stoichiometric and smooth KTaO$_3$ thin films at low substrate temperature. The SmScO$_3$ template serves three purposes (i) stabilizing the KTaO$_3$ surface by suppressing K evaporation from the KTaO$_3$ substrate surface at high temperature and high vacuum (*20*), (ii) inhibiting the migration of native defects from the KTaO$_3$ substrate to the film (*36*) that may deteriorate the superconductivity (fig. S8) and (iii) suppressing the transfer of charge carriers from the film to the substrate area (*37*) where more disorder is expected (*5*, *38*) (fig. S9). To alleviate the lattice parameter and oxygen octahedral rotation mismatch between SmScO$_3$ (space group: *Pbnm*) (*39*) and KTaO$_3$ (space group: $Pm\bar{3}m$) (*40*), we restricted the thickness of SmScO$_3$ template to be ≈1 nm. X-ray diffraction and atomic force microscopy validate this heterostructure design and indicate that the (pseudo-)homoepitaxial KTaO$_3$ thin film grown on SmScO$_3$-buffered KTaO$_3$ (111) substrate is stoichiometric and atomically smooth (fig. S7D-H).

We characterized the interfacial structures of the LaAlO$_3$/KTaO$_3$/SmScO$_3$/KTaO$_3$ (111) heterostructures using scanning transmission electron microscopy (STEM) (*31*). High-angle annular dark field (HAADF) cross-sectional images of the heterostructures (Fig. 2A, C and D) confirm atomically-sharp interfaces of SmScO$_3$/KTaO$_3$ (111) substrate (Fig. 2C) and LaAlO$_3$/KTaO$_3$ (111) thin film (Fig. 2D), which means that thin SmScO$_3$ template could protect the unstable (111) surface of the KTO substrate (*20*) under the highly reducing atmosphere of ≈$10^{-6}$ Torr at 973 K. The KTaO$_3$ thin film is fully epitaxial to the KTaO$_3$ substrate through the underlying SmScO$_3$ template without any misfit dislocations, facilitated by a small 2.6% lattice mismatch between SmScO$_3$ and KTaO$_3$ (111). The crystal structure of ≈1 nm-thick SmScO$_3$ film appears to adopt a (psuedo-)cubic structure, as can be seen in the alignment of Sm columns along the [001] direction without a buckling angle (fig. S10). This allows the KTaO$_3$ (111) thin film to be coherently grown with the desired cubic structure due to the (pseudo-)cubic SmScO$_3$ template. The STEM image of Fig. 2A shows a contrast variation near the LaAlO$_3$/KTaO$_3$ (111) thin film and SmScO$_3$/KTaO$_3$ (111) substrate interfaces. We attribute the brighter contrast in the KTaO$_3$ (111) thin film near the top interface of LaAlO$_3$/KTaO$_3$ (111) thin film to the presence of doubly charged oxygen vacancies ($V_O^{\cdot\cdot}$), which are predicted to increase the Ta-O and decrease the K-O bond lengths by ≈1.66% (*40*). This corresponds to expansion along the out-of-plane direction and increase of $c/a$ and cell volume (Fig. 2A and B). We attribute the brighter contrast in the KTaO$_3$ (111) substrate near the bottom SmScO$_3$/KTaO$_3$ (111) substrate interface to singly-charged K vacancies ($V_K'$), which have a lower formation energy of ≈0.06 eV than $V_O^{\cdot\cdot}$ (≈0.18 eV) and do not change the cell dimension (*40*). These observations directly confirm the higher crystalline qualities of the KTaO$_3$ (111) thin film compared to the KTaO$_3$ (111) bulk single crystal substrate.

**Electronic-grade epitaxial (111) KTaO$_3$ heterostructures**

These lower defect concentrations improve the normal state and superconducting state properties of the 2DEG at the LaAlO$_3$/KTaO$_3$ interface. We first compare normal-state transport measurements of two different heterostructures: LaAlO$_3$/KTaO$_3$ (111) substrate (denoted as "Bulk" and LaAlO$_3$/KTaO$_3$/SmScO$_3$/KTaO$_3$ (111) substrate (denoted as "Film") (Fig. 3) (*31*). As shown in Fig. 3a, 2DEGs are created at the LaAlO$_3$/KTaO$_3$ interface. The amorphous LaAlO$_3$ layer is grown *in situ* in the case of LaAlO$_3$/KTaO$_3$/SmScO$_3$/KTaO$_3$ (111) heterostructures to produce clean LaAlO$_3$/KTaO$_3$ interface (*41*). Normal state transport data (2 K < $T$ < 300 K) were obtained with a Van der Pauw geometry that probe the entire sample surface (Methods). Fig. 3B shows the $R_{sq}$-$T$ data of the "Bulk" and "Film" samples. The Film sample shows much lower $R_{sq}$ in the normal state (Fig. 3B) despite having nearly the same $n_{2D}$ as the Bulk sample at 10 K (Fig. 3C). We attribute the lower $R_{sq}$ to the high carrier mobility ($\mu$) of the Film sample (≈150 cm$^2$/Vs at 10 K) compared to the Bulk sample (≈48 cm$^2$/Vs at 10 K) (Fig. 3D). We tested multiple samples with the similar structures and summarize their properties at 10 K in a $\mu$-$n_{2D}$ diagram (Fig. 3E). This clearly demonstrates that the LaAlO$_3$/KTaO$_3$/SmScO$_3$/KTaO$_3$ (111) (red up-triangles, Fig. 3E) samples generally possess higher $\mu$ within the same $n_{2D}$ range compared to the LaAlO$_3$/KTaO$_3$ (111) samples (black down-triangles, Fig. 3E). We attribute the differences in $\mu$ to the lower point defect concentrations, which control the low temperature mobility. To further confirm the superior quality of the LaAlO$_3$/KTaO$_3$/SmScO$_3$/KTaO$_3$ (111) samples, we analyzed the $n_{2D}$ dependence of the mean free path ($l$) and Ioffe-Regel parameter $k_F l$ at 10 K, with $k_F$ the Fermi wave vector (fig. S11 and S12) (*31*). This confirms that the LaAlO$_3$/KTaO$_3$/SmScO$_3$/KTaO$_3$ (111) samples also possess higher $l$ and $k_F l$ than the LaAlO$_3$/KTaO$_3$ (111) samples and those reported in the literature (*4*, *5*) regardless of $n_{2D}$. Our analyses presented so far –X-ray diffraction, atomic force microscopy, STEM, and electrical transport measurements– cement hybrid PLD as a new synthesis method producing electronic-grade KTaO$_3$ thin films significantly cleaner than their bulk single crystal counterparts.

**Superconductivity**

We patterned Hall bars along the [11-2] and [1-10] on the Bulk and Film samples to investigate the superconductivity in KTaO$_3$, (Fig. 4A) (*31*). In the main text, we focus on electrical transport. Additional data on the Berezinskii-Kosterlitz-Thouless (BKT) transitions (fig. S13) and magnetotransport (fig. S14-S16) are provided. Fig. 4A shows the $R_{sq}$ vs. $T$ data at $T$ < 2 K along the [11-2]. The Film sample shows a $T_c$ ≈1.5 K, which is 25% higher than the $T_c$ ≈1.2 K exhibited by the Bulk sample. The *V-I* curves at $T = 0.5$ K (Fig. 4B) confirm the enhanced superconductivity in the Film sample with a critical current $I_c$ (≈12.3 µA), substantially larger than $I_c$ of the Bulk sample (≈3.9 µA) possessing a similar $n_{2D}$. The $R_{sq}$-*T*, *V-I* data (Fig. 4, fig. S17-S19) suggests that reduced disorder (*i.e.*, higher $\mu$) eliminates signatures of disorder-induced inhomogeneities (*42*) observed in the superconductivity of KTaO$_3$ (111) such as anisotropic transport (*4*), residual resistance (*5*), and step-like *V-I* curves (fig. S18 and S19) which appear to be highly sensitive to $\mu$ (*5*).

We have demonstrated superior low temperature mobility and superconducting properties in 2DEGs formed by electronic-grade KTaO$_3$ heterostructures. The ability of the hybrid PLD to synthesize electronic-grade thin films made of compounds containing elements with very large vapor pressure mismatch also opens the door to exploring the untapped potential of many unexplored thin film systems and derivative heterostructures such as freestanding membranes (*43*). Beyond KTaO$_3$, a synthetic method of compounds containing both highly volatile and refractory elements in thin film forms holds promise for revolutionizing a wide range of technologies, including environment-friendly microelectromechanical systems (*44*), integrated solid-state batteries (*45*), and emerging quantum technologies (*19*, *46*).

**Acknowledgments:**

**Funding:**

Gordon and Betty Moore Foundation's EPiQS Initiative, grant GBMF9065 (CBE)

Vannevar Bush Faculty Fellowship (N00014-20-1-2844) (CBE)

ONR MURI (N00014-21-1-2537) (JL)

DOE-QIS program award number DE-SC0022277 (MY, RR, JL)

US Department of Energy, Office of Science, Office of Basic Energy Sciences, award number DE-FG02-06ER46327

**Author contributions:**
J.K., J.W.L., and C.B.E. conceived the project. C.B.E., J.L., M.S.R., S.Y.C., Z.K.L., and S.H.O. supervised the project. J.K., J.W.L., and K.E. fabricated and characterized the thin-film samples. S.S. and Z.K.L. carried out the thermodynamic calculations. J.S., G.Y.K, S.Y.C., and S.H.O. performed the STEM measurements. M.Y., P.P., and N.C. carried out the electrical transport measurements. M.Y. and R.R. carried out the magnetotransport and superconductivity measurements. J.K., M.Y., J.W.L., S.S., S.Y.C., Z.K.L., J.L., and C.B.E. prepared the manuscript. C.B.E. directed the overall research.

**Competing interests:** Authors declare that they have no competing interests.

**Data and materials availability:** All data are available in the main text or the supplementary materials.

**Supplementary Materials**

Materials and Methods

Supplementary Text

Figs. S1 to S20

Tables S1 and S2

References (*47–51*)

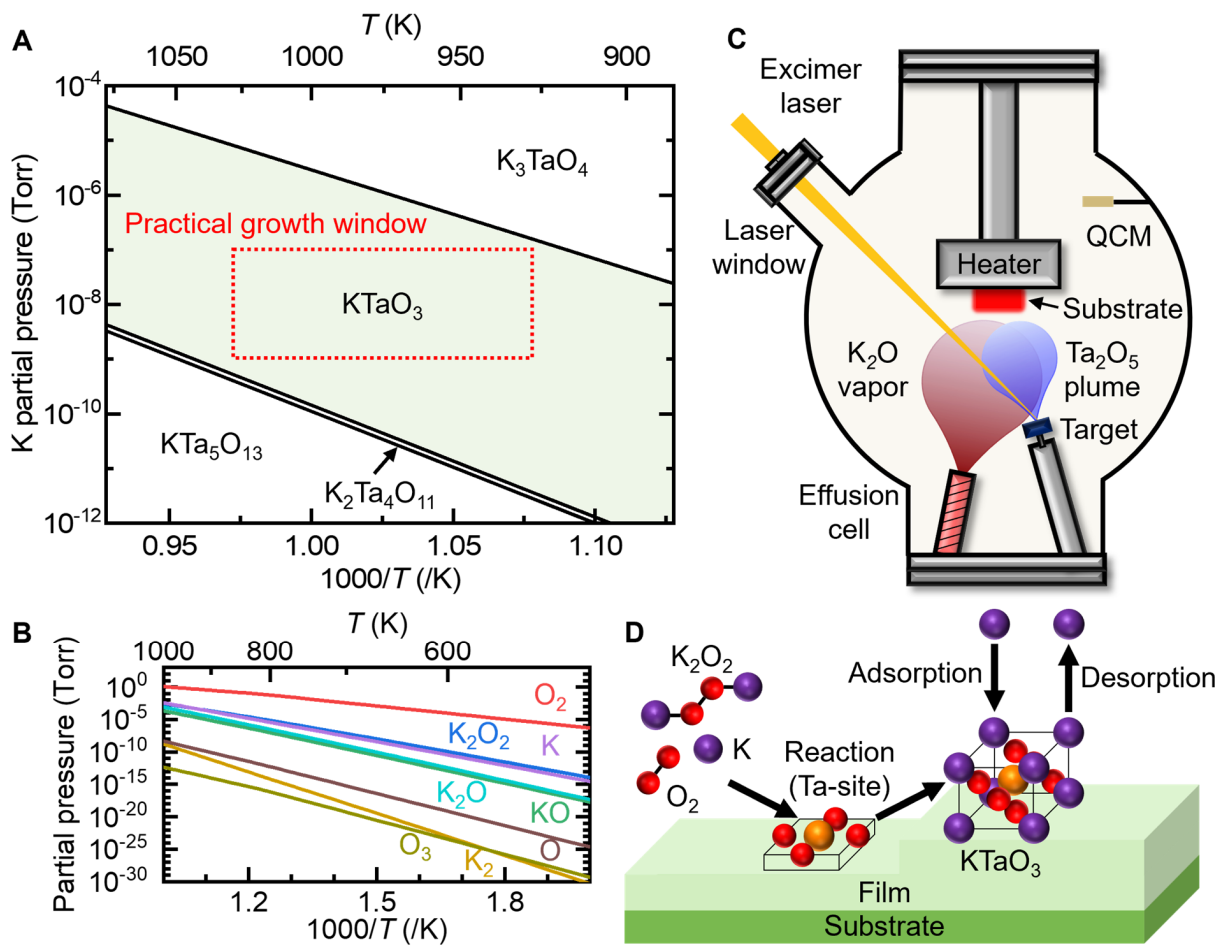

**Fig. 1. Thermodynamics-guided epitaxial growth of KTaO₃ thin films by hybrid PLD.** (**A**) Phase region of KTaO₃ as a function of K partial pressure and temperature at fixed O₂ partial pressure of $10^{-6}$ Torr. Practical experimental parameters within the KTaO₃ growth window are marked by a red box. (**B**) Vapor pressure of gas species in the K-O system. (**C**) Schematic illustrating the hybrid PLD method for KTaO₃ thin film growth. QCM: quartz crystal microbalance. (**D**) Schematic illustrating the adsorption-controlled growth of KTaO₃ thin films.

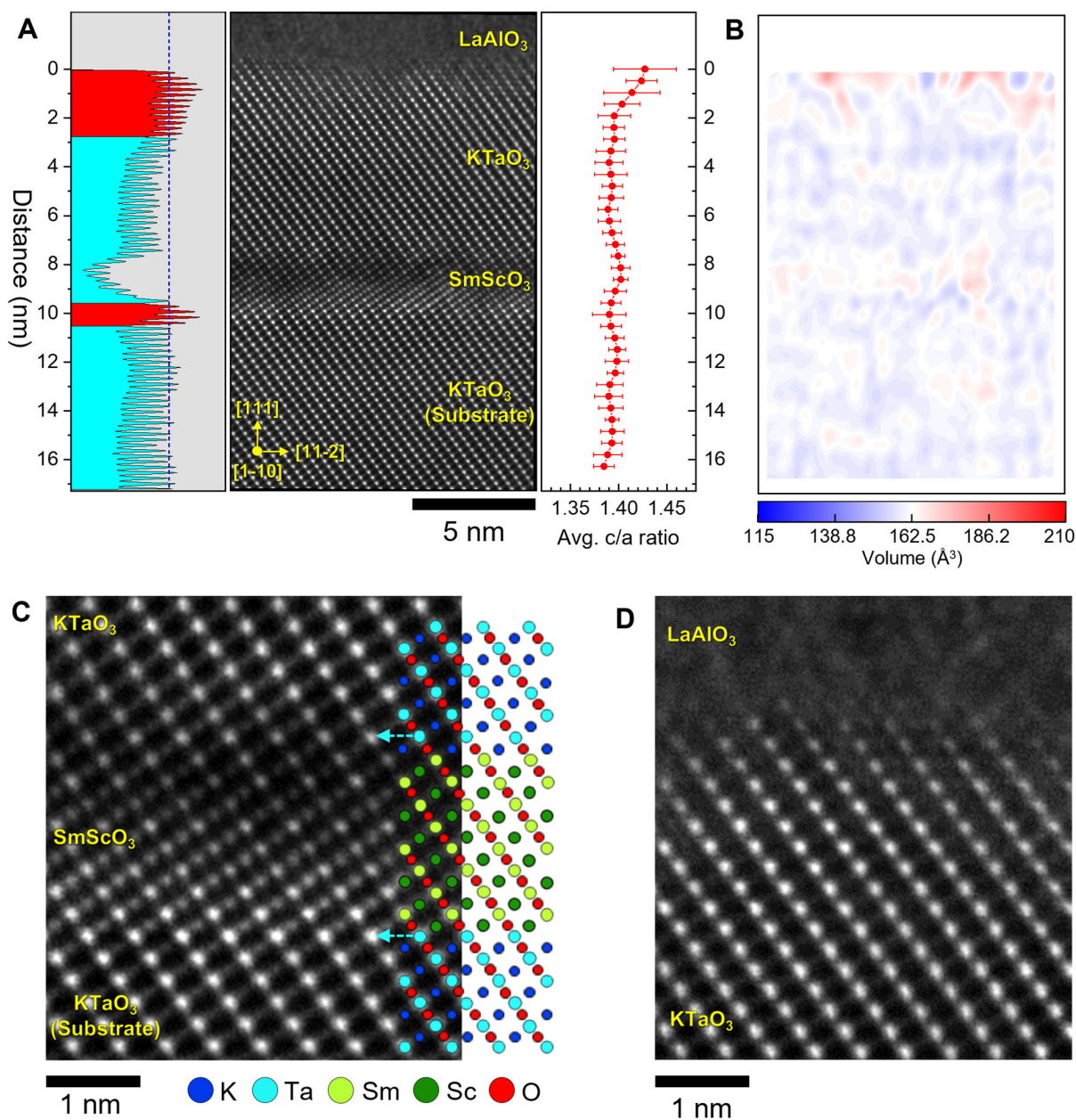

**Fig. 2. Atomically sharp interfaces in epitaxial LaAlO$_3$/KTaO$_3$/SmScO$_3$/KTaO$_3$ (111) heterostructure.** (**A**) HAADF-STEM image of cross-section viewed along the [1-10] and intensity profile. Red color in the intensity profile indicates the regions appearing brighter in the image near the interfaces of LaAlO$_3$/KTaO$_3$ thin film and SmScO$_3$/KTaO$_3$ substrate. (**B**) Volume color mapping and average volume measurement in (A). The average volume is close to the ideal volume in a cubic structure. Approximately ≈2 nm-thick regions of higher volume are observed near the LaAlO$_3$/KTaO$_3$ thin film interface, indicating the oxygen vacancies are confined to the top of the LaAlO$_3$/KTaO$_3$ thin film interface. (**C**) Magnified image of the KTaO$_3$ thin film/SmScO$_3$/KTaO$_3$ substrate region. (**D**) Magnified image of the LaAlO$_3$/KTaO$_3$ thin film region.

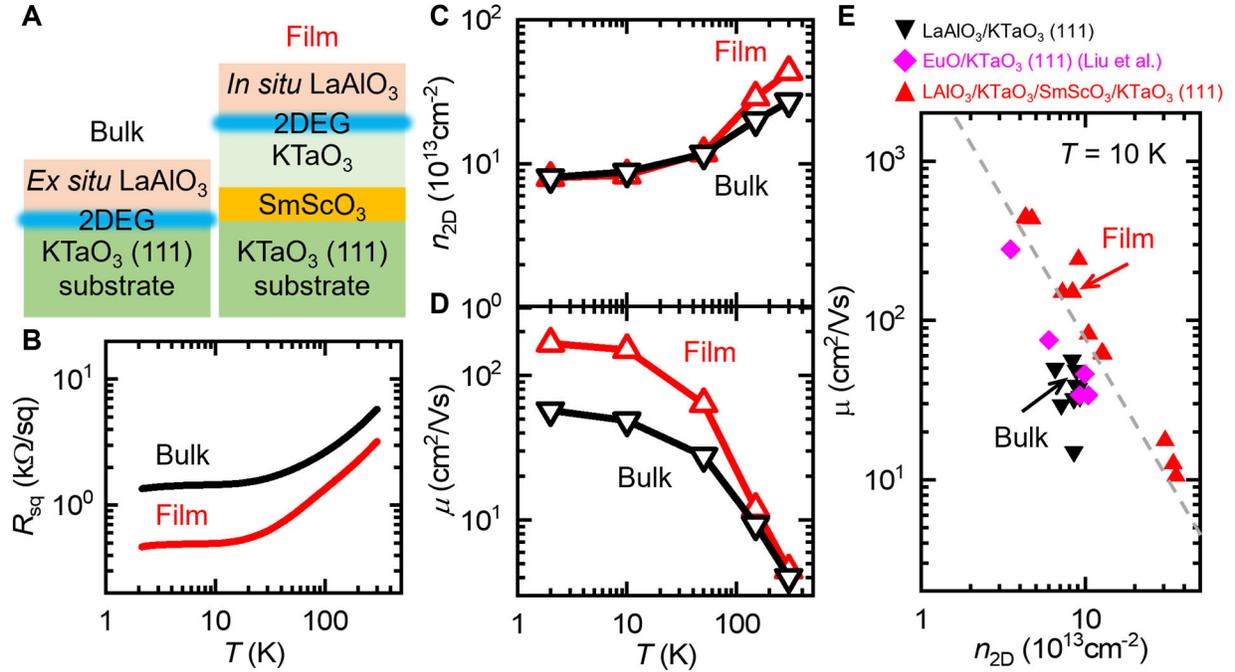

**Fig. 3. Electrical transport measurements of KTaO₃ (111).** (**A**) Schematic illustrating the structures of the measured samples. In the bulk case, the LaAlO₃ overlayer is always grown *ex situ*, which inevitably creates the 2DEG in the first few nm of KTaO₃ with high defect density. In the case of KTaO₃ thin film, the LaAlO₃ layer is grown *in situ* and the surface of KTaO₃ has low defect density, which results in enhanced $\mu$. (**B-D**), Temperature dependence (2-300 K) of (B) $R_{sq}$, (C) $n_{2D}$, (D) $\mu$ of LaAlO₃/KTaO₃ (111) (Bulk) and LaAlO₃/KTaO₃/SmScO₃/KTaO₃ (111) (Film) heterostructures. The measurements in (B-D) are performed in a Van der Pauw geometry. (**E**) Distribution of $\mu$ and $n_{2D}$ estimated from Hall measurements at $T$ = 10 K. The samples shown in (B-D) marked with arrows. Purple diamonds are data at $T$ = 10 K from different growth conditions of EuO from Liu et al (*4*).

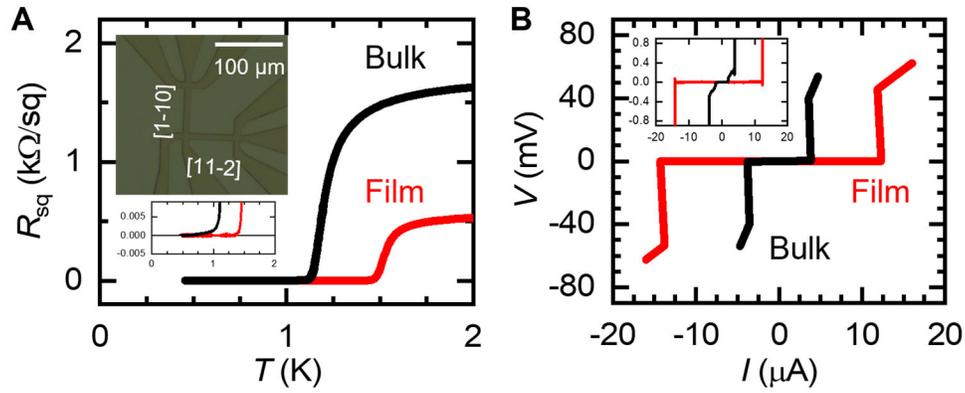

**Fig. 4. Superconductivity of KTaO₃ (111).** (**A**) Temperature dependence of $R_{sq}$ along the [11-2] on Hall bars. The insets show an optical micrograph of Hall bars (top) and a magnified view near the transition (bottom). (**B**) *V-I* curves along the [11-2] measured at $T$ = 0.5 K on Hall bars. The inset shows a magnified view near the transition.

**Materials and Methods**

Thermodynamic calculations of the K-Ta-O system

Thermodynamic database used in the present work is based on the SGTE substance database (i.e., the SSUB5) (*47*). However, thermodynamic properties for the six ternary oxides ($KTa_5O_{13}$, $K_2Ta_4O_{11}$, $KTaO_3$, $K_3TaO_4$, $K_3Ta_8O_{21}$, and $K_2Ta_{15}O_{32}$) are absent in SSUB5. In the present work, their thermodynamic properties are estimated with the reference states being the binary oxides (and pure element Ta) in SSUB5 according to the reactions in Supplementary Table 1. The reaction enthalpies ΔH for these ternary oxides are based on density functional theory (DFT) based first-principles calculations in the literature, i.e., the Materials Project (MP) (*48*) and the Open Quantum Materials Database (OQMD) (*49*); see the predicted ΔH values in Supplementary Table 1. Using the presently generated K-Ta-O database (SSUB5 together with the DFT-based ΔH values), we can perform thermodynamic calculations using the Thermo-Calc software (*50*).

Film synthesis and structural characterizations

Hybrid PLD using a KrF excimer laser (248 nm, LPX 300, Coherent) and a home-built effusion cell were used to grow epitaxial $KTaO_3$ thin films on single crystal $KTaO_3$ (111), $SrTiO_3$ (111), $SrTiO_3$ (001) substrates. To obtain the $TiO_2$-terminated substrates, as-received $SrTiO_3$ substrates were etched using buffered hydrofluoric acid (Sigma Aldrich) for 1 min and annealed at 1273 K for 6 hr. The $KTaO_3$ growth was carried out at a heater temperature of 973 K in a chamber pressure of $10^{-6}$ Torr. To supply tantalum, a ceramic target of $Ta_2O_5$ (purity 99.9%, Praxair) was ablated at using a laser fluence of 0.5 J/cm$^2$ and a laser repetition rate of 20 Hz with a target-to-substrate distance of 65 mm. To supply potassium, 5 g of $K_2O$ powders (purity 99.9%, Nanoshel) were used as source and loaded in a magnesium oxide crucible of cylindrical shape with a diameter of 16 mm and a length of 100 mm. To protect the effusion cell from reaction with potassium vapors, the effusion cell was shielded with tantalum foils. The crucible was heated using a home-built effusion cell and $K_2O$ source was pre-evaporated at an effusion cell temperature of 873 K in vacuum for de-gassing. Following pre-evaporation, the effusion cell temperature was lowered to 750 K and maintained during the $KTaO_3$ growth. Following the $KTaO_3$ growth, the samples were cooled down to 823 K by quenching in a static oxygen pressure of 500 Torr and annealed *in situ* for 30 min. The growth rate for $KTaO_3$ was approximately 0.0044 Å per laser pulse (≈2.5 u.c./min). The laser fluence was kept at slightly above the ablation threshold for $Ta_2O_5$ ceramic to minimize the high kinetic energy ion bombardment during the $KTaO_3$ growth. The $SmScO_3$ growth was carried out by conventional PLD at a heater temperature of 973 K in a dynamic oxygen pressure of 20 mTorr with a laser fluence of 0.9 J/cm$^2$ and a laser repetition rate of 1 Hz from a ceramic target (Praxair) of the same composition with a target-to-substrate distance of 65 mm. The growth rate of $SmScO_3$ was approximately 0.066 Å per laser pulse. Following the growth, the samples were cooled to room temperature by quenching in a static oxygen pressure of 500 Torr. The $LaAlO_3$ growth was carried out *in situ* by conventional PLD at a heater temperature of 673 K in a dynamic oxygen pressure of $10^{-5}$ Torr with a laser fluence of 1.6 J/cm$^2$ and a laser repetition rate of 1 Hz from a single crystal $LaAlO_3$ target (Crystec) with a target-to-substrate distance of 65 mm. The growth rate of $LaAlO_3$ was approximately 0.11 Å per laser pulse. Following the growth, the samples were cooled to room temperature by quenching in the growth atmosphere. X-ray diffraction measurements were conducted with a high-resolution X-ray diffractometer (Bruker). Surface topography images were collected with an atomic force microscope operated in a tapping mode (Veeco).

Scanning transmission electron microscopy
The cross-sectional STEM sample having [1-10] projection was prepared using a dual-beam focused ion beam system (Helios G3, FEI) to observe the interfacial structure. We used a Ga ion beam at 30 kV to make a thin specimen and then used different acceleration voltages from 5 to 1 kV for the sample cleaning process to reduce the Ga damage. The atomic structure observation was performed using a STEM (JEM-ARM200F, JEOL, Japan) at 200 kV equipped with a fifth-order probe corrector (ASCOR, CEOS GmbH, Germany) at Materials Imaging & Analysis Center of POSTECH in South Korea. The optimum size of the electron probe for STEM observation was ~78 pm. The collection semi-angles of the HAADF detector were adjusted from 68 to 280 mrad to collect scattered electrons in a large angle for clear Z-sensitive images. HAADF-STEM images were acquired using Smart Align (HREM Research Inc., Japan), which conducted the multi-stack of images and aligned them using rigid registration to correct the sample drift and scan distortions. The obtained raw images were processed using a bandpass Difference filter with a local window to reduce background noise (Filters Pro, HREM Research Inc., Japan). HAADF-STEM image analysis was performed by Python with the customized atomic analysis code. All atomic coordinates were determined by the centroid of each atomic column. $c/a$ ratio and volume mapping were conducted by the nearest Ta sites of $KTaO_3$ or Sm sites of $SmScO_3$ along the [111] of $c$ and [11-2] of $a$ directions.

Device fabrication
Hall bars were fabricated with standard photolithography. First, Au markers were deposited for alignment purpose. AZ4210 photoresist was patterned to cover and protect the Hall bar-shaped regions. The exposed regions underwent inductively couple plasma reactive ion etching (ICPRIE) for 18 minutes in a Plasma-Therm APEX ICPRIE. During the etching, 5 sccm of $BCl_3$, 50 sccm of $Cl_2$ and 5 sccm of Ar were used as gas etchant, with RIE power set to 100 W. After the etching, photoresist was removed in MICROPOSIT Remover 1165, then acetone and IPA. The etching depth is measured to be ≈50 nm by atomic force microscope.

Electrical transport
The electrical transport measurements were carried out using a four-contact van der Pauw geometry over a temperature range of 2 to 300 K. Two van der Pauw sheet resistance and two Hall measurement configurations were switched between while sweeping a magnetic field over a range of -1 to 1 T. Individual configuration resistances were computed by sourcing a DC current of alternating polarity and combining the voltages to compute a resistance free of voltage noise from effects including the Seebeck effect. The two resistances from the sheet-resistance configurations ($R_1$, $R_2$) were combined using the van der Pauw equation, $1 = \exp(-\pi R_1/R_{sq}) + \exp(-\pi R_2/R_{sq})$, to determine the sheet resistance, $R_{sq}$. The two Hall resistances were averaged to compute the Hall resistance, $R_H$. The carrier density, $n_{2D}$, was computed with the equation $n_{2D} = 1/[(dR_H/dB)q]$, where I is the magnitude of the d.c. current sourced, and q the electron charge. The mobility, μ, was computed with the equation $\mu = 1/(R_{sq} n_{2D} q)$. $k_F$ and $l$ were computed with the equations $k_F = (2\pi n_s)^{1/2}$ and $l = h/(e^2 k_F R_s)$.

Magnetotransport and Superconductivity
Superconductivity measurements on Hall bars below 2 K were carried out in a Quantum Design PPMS refrigerator with a dilution unit. Source voltages were applied by a 24-bit digital/analog converter National Instruments PXI-4461, which can also simultaneously perform 24-bit analog/digital conversion. Current biasing was achieved by shunting the Hall bar device with a 300 kΩ in-series resistor. The drain current and voltages were measured after amplification by a Krohn-Hite Model 7008 Multi-channel Pre-amplifier. For each current-voltage characteristic, the current sweep started from 0 bias, then went to maximum bias, to minimum bias, and finally finished at 0 bias. The whole sweep took ≈20 s. For 0 bias resistance measurement, a small AC current $I = I_0 \cos(2\pi f t)$ was sourced where $f = 13$ Hz and $I_0 < 100$ nA. The voltage $V_0$ across the Hallbar channel was measured by a homemade digital lock-in amplifier to get four-terminal resistance $R = V_0 / I_0$. High-field magnetoresistance measurements on Hall bars were performed in a Leiden MNK dilution fridge equipped with an Oxford 18T magnet.

**Supplementary Text**

Thermodynamic calculations of $KTaO_3$ phase diagram

Thermodynamic database used in the present work is based on the SGTE substance database (i.e., the SSUB5) (47). However, thermodynamic properties for the six ternary oxides ($KTa_5O_{13}$, $K_2Ta_4O_{11}$, $KTaO_3$, $K_3TaO_4$, $K_3Ta_8O_{21}$, and $K_2Ta_{15}O_{32}$) are absent in SSUB5. In the present work, their thermodynamic properties are estimated with the reference states being the binary oxides (and pure element Ta) in SSUB5 according to the reactions in Supplementary Table 1. The reaction enthalpies ΔH for these ternary oxides are based on density functional theory (DFT) based first-principles calculations in the literature, i.e., the Materials Project (MP) (48) and the Open Quantum Materials Database (OQMD) (49); see the predicted ΔH values in table S1. Using the presently generated K-Ta-O database (SSUB5 together with the DFT-based ΔH values), we can perform thermodynamic calculations using the Thermo-Calc software (50).

Synthesis of epitaxial $KTaO_3$ heterostructures and structural characterizations

Epitaxial $KTaO_3$ thin films of approximately 8 nm thickness were initially grown on single-terminated $SrTiO_3$ (001) and (111) substrates to experimentally determine the growth window. As expected, the growth on (001) orientation is more easily achieved than (111) orientation. This is well-illustrated by dramatic differences in the surface morphology with substrate temperature (Fig. S5 and S6). Since the substrate temperature has the most dominant effects on the growth window and surface morphology, only the effect of the substrate temperature is discussed here but other growth parameters also impact the film quality. Due to the hybrid nature of hybrid PLD, the growth parameter space encompasses parameters for conventional MBE and PLD. Common parameters include substrate temperature, background pressure, oxygen partial pressure. Parameters specific for MBE include source geometry and source temperature. Parameters specific for PLD include laser fluence, laser frequency, laser mask geometry, target-to-substrate distance. For hybrid PLD to work, the background pressure should be low to allow molecular flow of thermally evaporated sources, typically below $10^{-3}$ Torr. In this work, the background pressure during the growth was kept at ~$10^{-5}$ Torr. Such low-pressure condition may make the films susceptible to bombardment-induced defects from PLD if high laser fluence is employed. For this reason, laser fluence just above the ablation threshold was used to minimize the bombardment-induced defects.

Despite the polar nature of $KTaO_3$ (001) surface, we were able to achieve atomically-flat $KTaO_3$ films grown on $SrTiO_3$ (001) substrates over a wide range of growth parameters. Since the crystalline quality shows substantial improvements with increasing temperature, it is clearly advantageous to use high substrate temperature to obtain electronic-grade thin films (Fig. S5). In contrast, the (111) surface is highly susceptible to temperature-induced surface roughening at or above 1023 K, although the crystallinity continues to improve with increasing temperature (Fig. S6).

Due to the volatility of potassium, exposure to high temperature and high vacuum conditions is expected to severely degrade the surface quality of $KTaO_3$ substrate, compromising the homoepitaxial growth. To protect the $KTaO_3$ surface, we optimized the growth of ≈8 nm thick $SmScO_3$ thin films by conventional PLD and confirmed the high crystallinity and atomically-flat surface. To alleviate the lattice parameter and oxygen octahedral rotation mismatch between $SmScO_3$ (space group: *Pbnm*) (*39*) and $KTaO_3$ (space group: $Pm\bar{3}m$) (*40*), we restricted the thickness of $SmScO_3$ template to be ≈1 nm. We confirmed that the surfaces of ≈1 nm $SmScO_3$ and ≈8 nm $KTaO_3$ are atomically-flat with no formation of any extended defects (Fig. S7).

Transport properties of $LaAlO_3/KTaO_3/KTaO_3$ heterostructures
As illustrated in Fig. S7, the crystalline qualities of homoepitaxial $KTaO_3$ films ($KTaO_3/KTaO_3$ (111)) are compromised without a $SmScO_3$ template and as a result we observe semiconducting ($dR/dT > 0$ at low temperatures) behavior below 15 K (Fig. S8A). For comparison, the data for homoepitaxial ($LaAlO_3/KTaO_3/KTaO_3$ (111)) sample are presented with the data for Film_4 ($LaAlO_3/KTaO_3/SmScO_3/KTaO_3$ (111), see the description in the main text on Fig. 3 for details) sample. The two heterostructures are sister samples grown in the same deposition, suggesting the differences are not due to growth-to-growth variations. We have also tested the homoepitaxial sample for superconductivity in the 4-terminal geometry and the homoepitaxial sample shows a highly insulating behavior ($R_{4T}$ ≈50kΩ across 5 mm) and does not show superconductivity down to 50 mK (Fig. S8D).

Transport properties of $LaAlO_3/SmScO_3/KTaO_3$ heterostructures
Here, we present a control experiment on $LaAlO_3/SmScO_3/KTaO_3$ (111) heterostructures to investigate the contribution of $KTaO_3$ substrate. To estimate the upper limit of $n_{2D}$ that may be doped across the insulating $SmScO_3$ template, we grew $LaAlO_3$ directly on ≈1 nm-thick $SmScO_3$ ("Buffered bulk" in Fig. S9B). We observe that the $n_{2D}$ is ≈1.7x10$^{13}$cm$^{-2}$ in such heterostructures. Despite the non-zero $n_{2D}$, this heterostructure is not superconducting (Fig. S9D) and has much higher $R_{sq}$ than both Bulk_7 and Film_4 samples (Fig. S9A, see Fig. S9E for the labels), likely due to the enhanced disorder underneath the $SmScO_3$ in the $KTaO_3$ substate. In addition, the thickness of the $KTaO_3$ film (≈8 nm) is much larger than the estimated superconducting layer thickness (≈4 nm) (*5*). Therefore, we conclude that the substrate does not contribute to superconductivity in the Film_4 sample, at least above 50 mK, which is much lower than the $T_{c,\rho=0}$ of the Film_4 sample (≈1.35 K).

Cleanliness of various heterostructures estimated from Van der Pauw method
In the literature, two common approaches to characterize and compare the cleanliness of superconductors are 1) ratio of mean free path (*l*) to Ginzburg-Landau coherence length ($\xi_{GL}$) and 2) calculating the product of mean free path (*l*) and Fermi wave vector ($k_F$). The mean free path is estimated from $l = h/(e^2 k_F R_{sq})$, where $h$ is the Planck constant, e is the electron charge, $R_{sq}$ is the sheet resistance. We estimate the cleanliness of the samples based on $R_{sq}$ at 10 K, which is well above the superconductivity onset temperature ($T_c^{onset}$ ≤3 K), but estimation at other temperatures show a similar trend. For full comparison, *l*, $k_F l$, and $\mu$ are compared in the $n_{2D}$ range of 1 to 50cm$^-$

$^2$. All values are at 10 K (except for green data on $V_G$ dependence from Chen et al. (*5*), which is at 8 K). Notably, it is possible to grow KTaO$_3$ thin films (LaAlO$_3$/KTaO$_3$/SmScO$_3$/KTaO$_3$ (111)) cleaner than LaAlO$_3$/KTaO$_3$ (111) substate samples for a range of $n_{2D}$ which are produced by adjusting only the LaAlO$_3$ growth atmosphere (from 10$^{-5}$ Torr oxygen for low $n_{2D}$ to 10$^{-5}$ Torr argon for high $n_{2D}$). For a meaning comparison, we have grown multiple LaAlO$_3$/KTaO$_3$ (111) samples using the same LaAlO$_3$ growth conditions (black down-triangles in Supplementary Fig. 11) as the thin film samples. None of the in-house bulk samples had higher cleanliness than the high-quality thin films (red up-triangles in Fig. S11). We also overlaid data on EuO/KTaO$_3$ (111) from Liu et al. (*4*) and LaAlO$_3$/KTaO$_3$ (111) from Chen et al. (*5*), which show that the KTaO$_3$ thin films exhibit much higher cleanliness; for example, $k_Fl$ for $n_{2D}$ in the range of 8-10x10$^{13}$cm$^{-2}$ are 51.9, 19.0, and 17.8 for thin film, EuO/KTaO$_3$ (111), and LaAlO$_3$/KTaO$_3$ (111) with +150 V back-gate voltage (positive bias leads to cleaner conditions, Fig. S11, see Chen et al. (*5*) for details).

Moreover, the high cleanliness of KTaO$_3$ thin films allow very high $n_{2D}$ ≈30.5x10$^{13}$cm$^{-2}$ to be realized at 10 K (52.2x10$^{13}$cm$^{-2}$ at 300 K). This is more than three times higher than typically achievable $n_{2D}$ ≈10-11x10$^{13}$cm$^{-2}$ in bulk samples (LaAlO$_3$/KTaO$_3$ (111)) at 10 K regardless of the doping method used. In our experience, typical values for n$_{2D}$ at 300 K and carrier loss for bulk samples when temperature is decreased from 300 K to 2 K are ≤24-30x10$^{13}$cm$^{-2}$ and about two thirds of the n$_{2D}$ at 300 K (≈16-20x10$^{13}$cm$^{-2}$) for $n_{2D}$ at 10 K ≈8x10$^{13}$cm$^{-2}$. We speculate that the large $n_{2D}$ achievable at 10 K (and also at 300 K) is attributable to two factors. First, thin films have a lower concentration of point defects, which act as charge trapping sites and impurity scattering centers, than bulk samples. The most probable defects responsible for such trends are potassium vacancies ($V_K$), which are nominally +1 charged but most stable in the -1 charge state, forming $V'_K$. Since potassium vacancies also have the lowest formation energies, it is plausible that the reduction of potassium non-stoichiometry in KTaO$_3$ is crucial to realizing electronic grade quality KTaO$_3$ with a high n$_{2D}$ that we observe in our KTaO$_3$ thin films. Second, the growths of LaAlO$_3$ overlayers were achieved *in situ* during the growth of KTaO$_3$ films without breaking the vacuum. The *in situ* growth may preserve the native surface state during the KTaO$_3$ growth since it does not expose the KTaO$_3$ surface to ambient atmosphere and moisture, which are known to result in surface reconstructions in the KTaO$_3$ (001) surface (*41*) (currently, details are unknown for the case of the KTaO$_3$ (111) surface).

Consistent with the above discussion, we observe that high $n_{2D}$ at 300 K is positively correlated with higher $\mu$ at low temperatures for similar $n_{2D}$ (e.g., 10 K, Fig. 3C, D). The temperature dependence data of $l$ and $k_Fl$ provide further evidence for cleanliness of thin films compared to bulk samples (Fig. S12). For this comparison, we performed a parallel analysis on a low $\mu$ bulk sample (LaAlO$_3$/KTaO$_3$ (111), "Bulk_4" in Fig. S12, also indicated by orange arrow in Fig. S11). The Bulk_4 sample was prepared by purposely inducing non-stoichiometry on the surface of KTaO$_3$ substrate. To achieve this, we annealed a KTaO$_3$ substrate at 973 K and a background pressure of 10$^{-5}$ Torr argon for 15 min. to induce potassium deficiency at the surface. To eliminate any oxygen vacancies that are created during this process, the substrate was subsequently annealed in 823 K and a static oxygen pressure of 500 Torr for 30 min, followed by an *in situ* LaAlO$_3$ growth using the same LaAlO$_3$ growth conditions as Bulk_7. We observe that the thin films possess superior cleanliness than bulk samples at all temperatures (Fig. S12B).

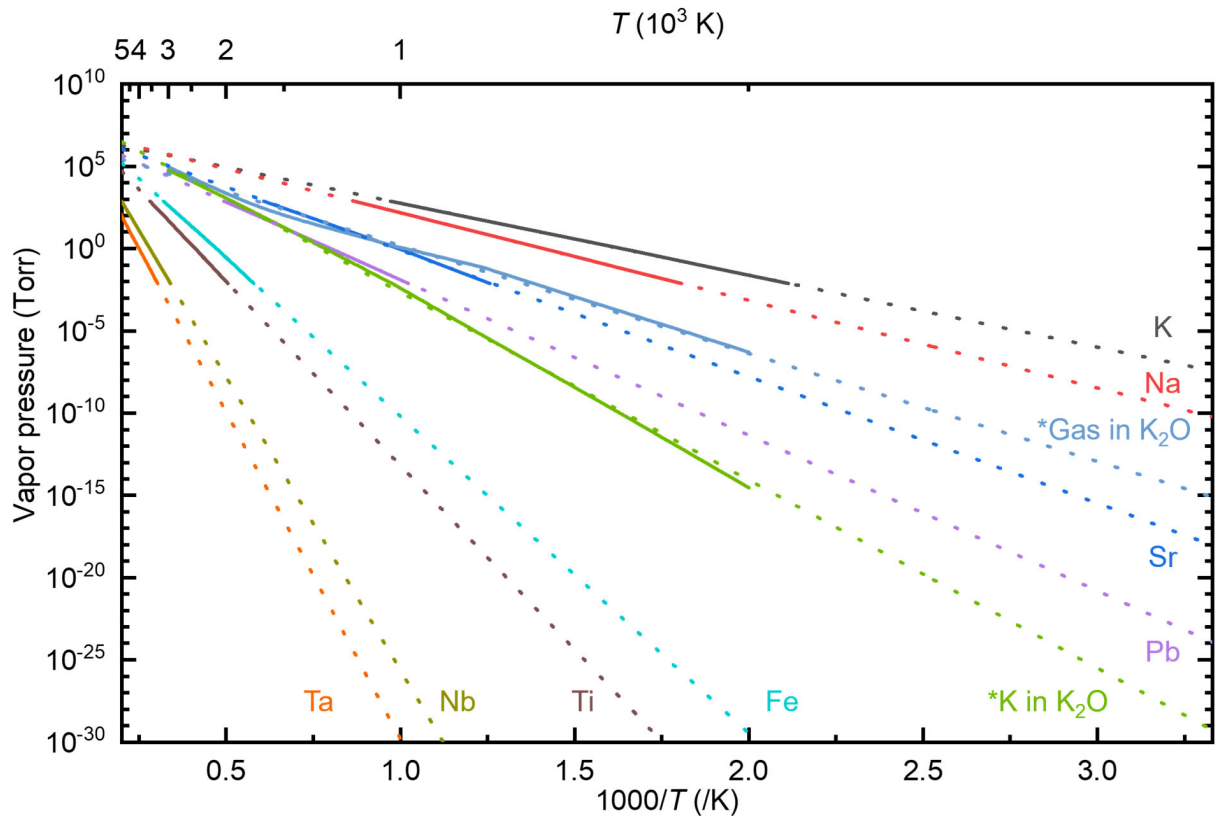

**Fig. S1. Vapor pressure of elemental metals.** Solid lines for elemental sources are obtained from the Trigg *et al* (*51*). *K and Gas in K$_2$O are the calculated partial vapor pressures of gas species shown in Fig. 1B. Dotted lines are fits to the solid lines and guides for the eye.

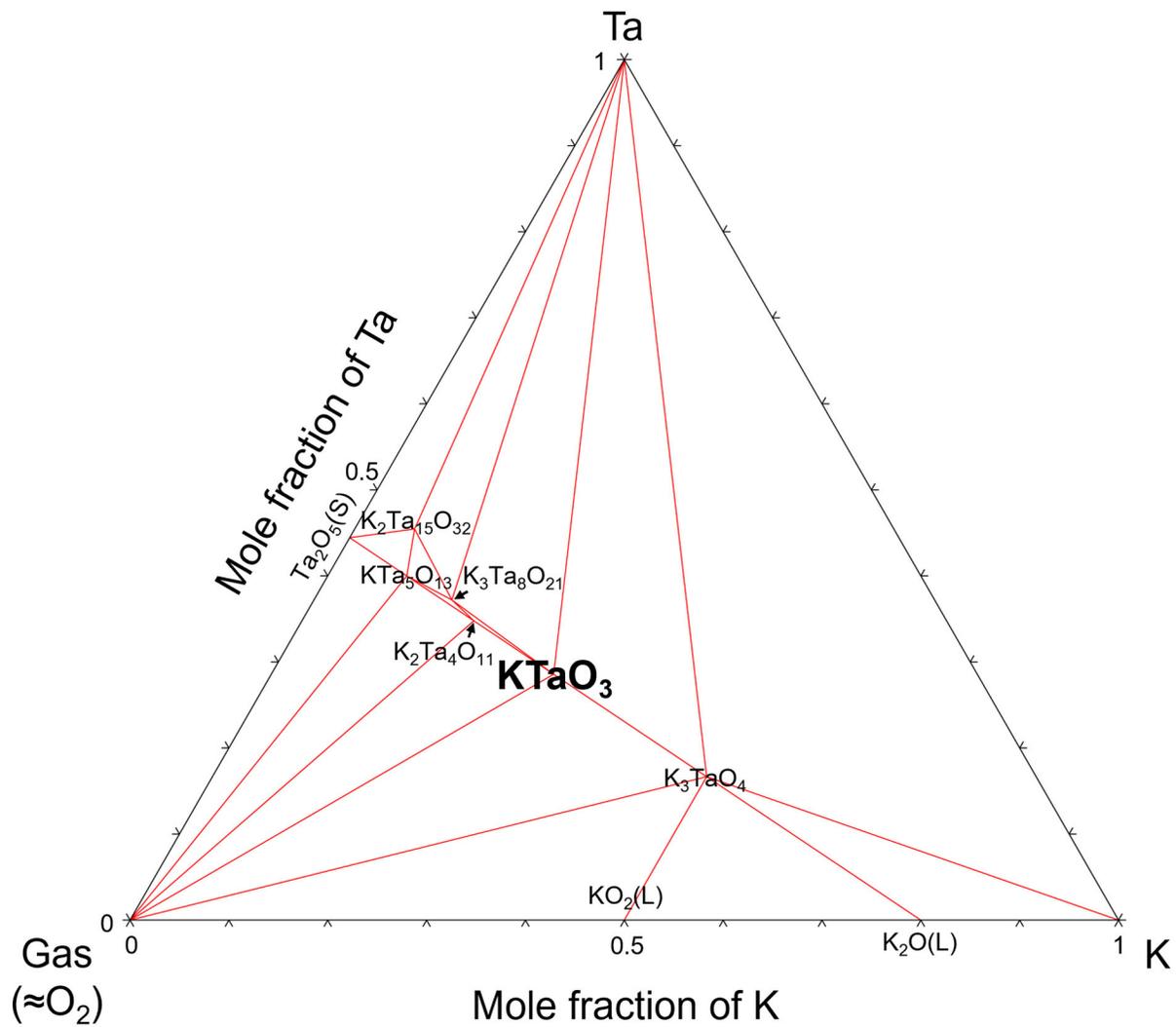

**Fig. S2. Isothermal section of K-Ta-O ternary phase diagram at** $T$ = **1000 K.**

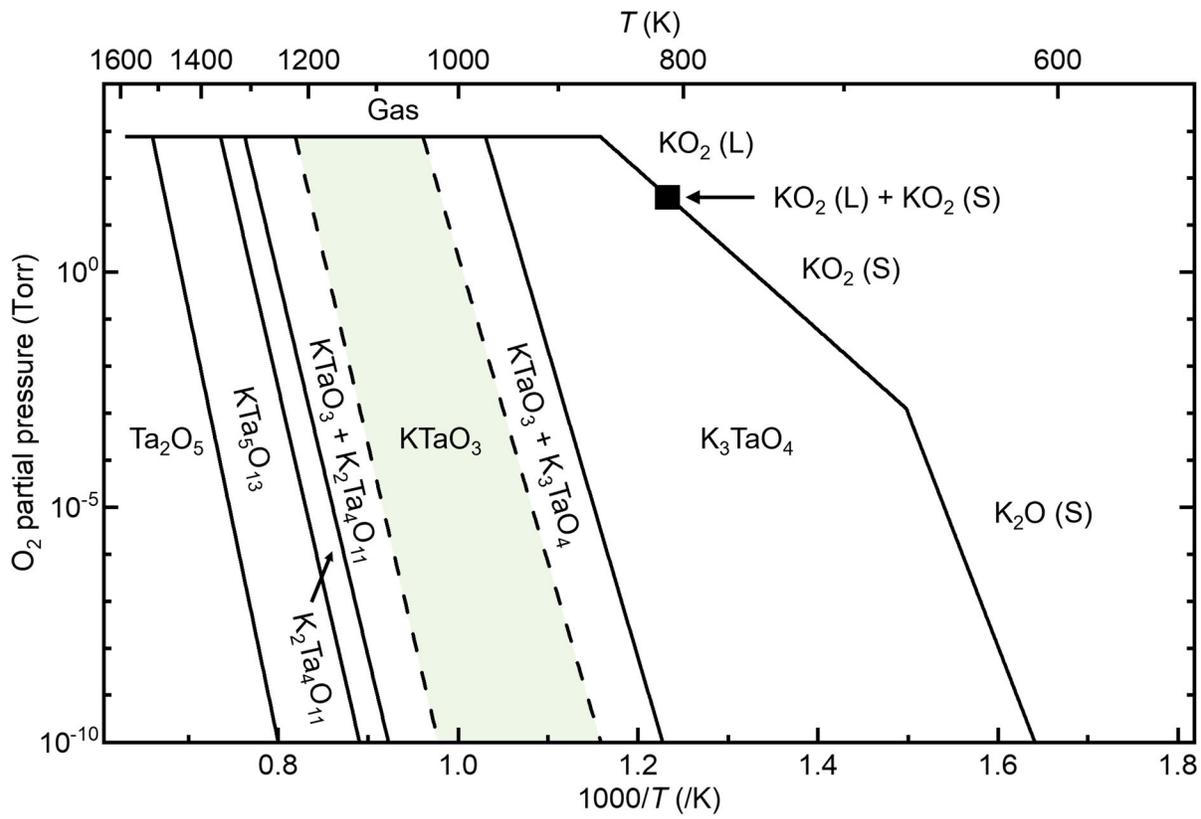

**Fig. S3.** Phase region of KTaO$_3$ as a function of O$_2$ partial pressure and temperature at fixed K partial pressure of 10$^{-8}$ Torr.

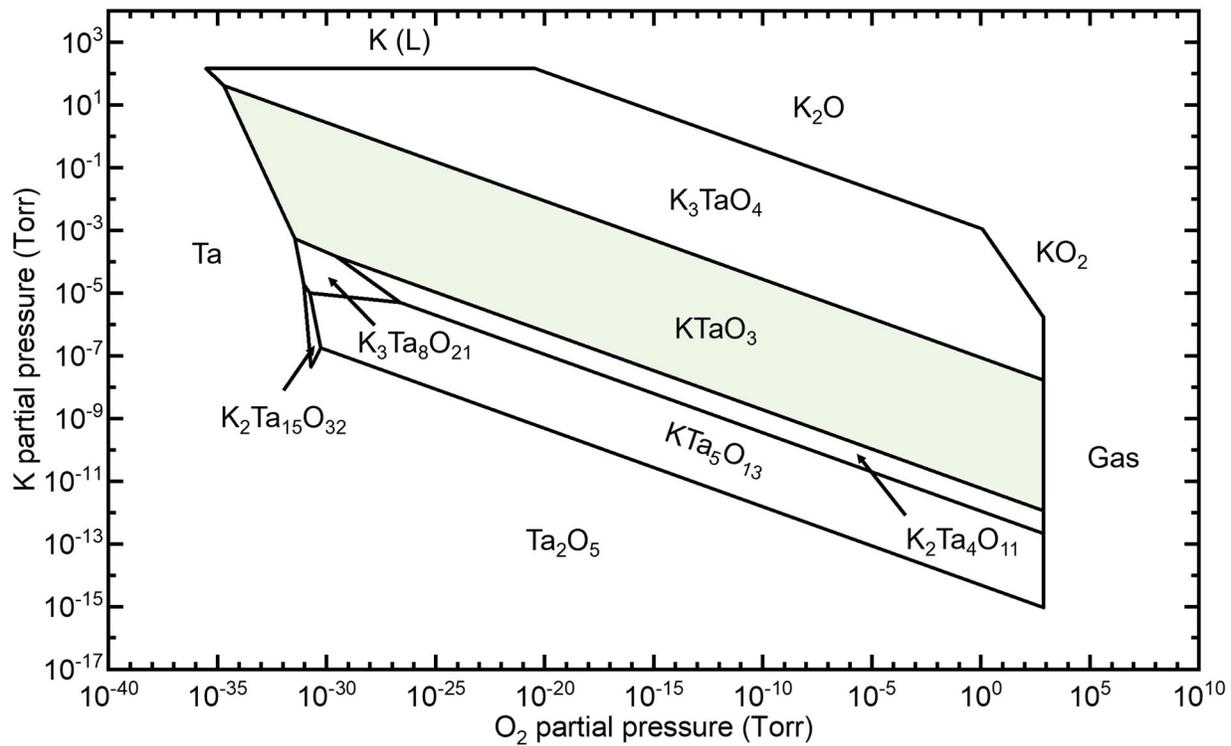

**Fig. S4. Phase region of KTaO$_3$ as a function of oxygen and potassium partial pressures at $T$ = 1000 K.**

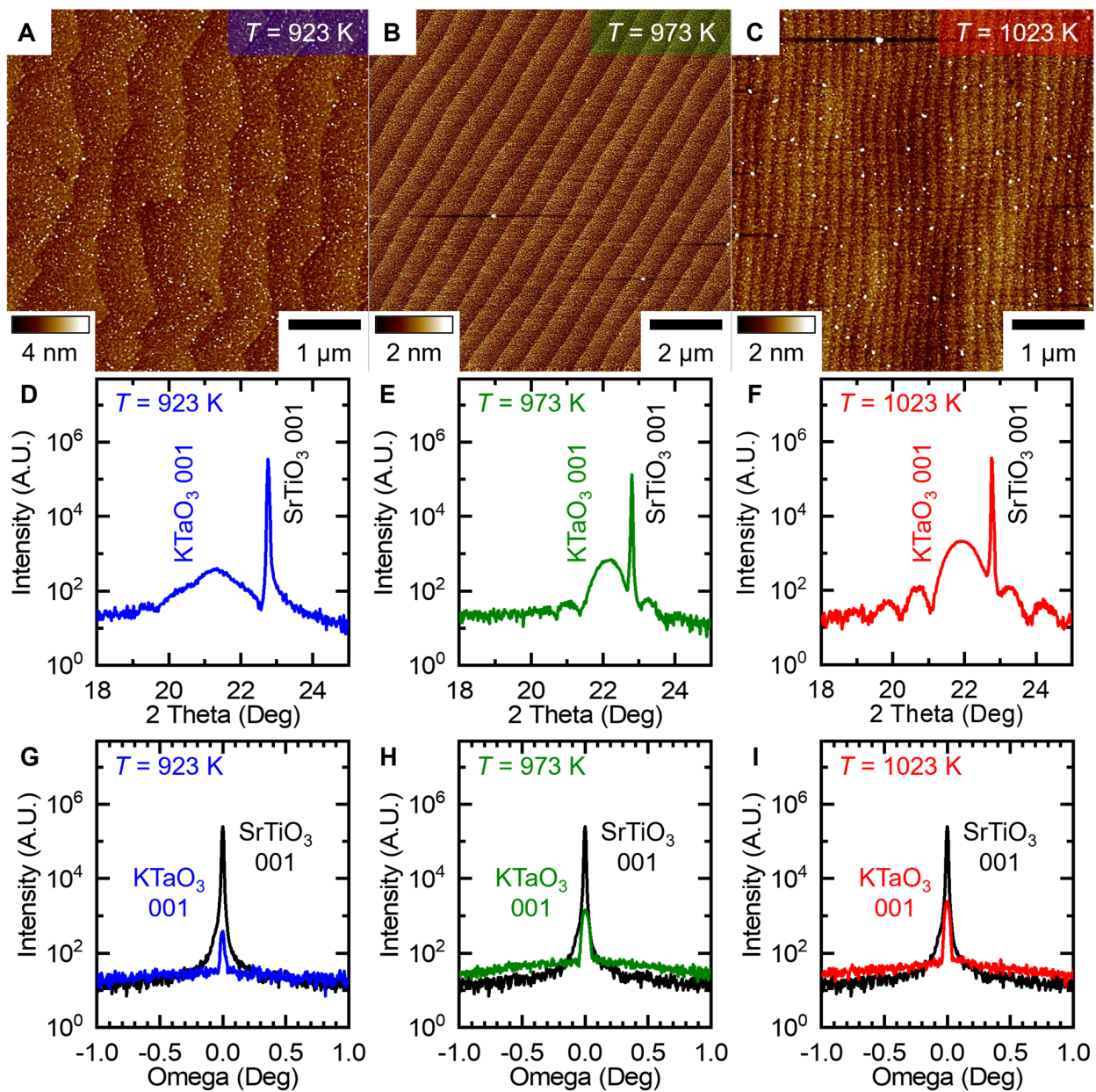

**Fig. S5. Structural characterizations of KTaO$_3$/SrTiO$_3$ (001) heterostructures.** (**A**-**C**) Surface topography of ≈8 nm KTaO$_3$/SrTiO$_3$ (001) heterostructures grown at a substrate temperature of (A) 923 K, (B) 973 K, (C) 1023 K. All heterostructures are grown with the same K$_2$O source temperatures of 750 K. In all conditions, (001) KTaO$_3$ can be synthesized with smooth surface morphologies. (**D**-**F**) θ-2θ X-ray diffraction line scans of ≈8 nm KTaO$_3$/SrTiO$_3$ (001) heterostructures grown at a substrate temperature of (D) 923 K, (E) 973 K, (F) 1023 K. (**G**-**I**) Rocking curves about the KTaO$_3$ 001-diffraction conditions of ≈8 nm KTaO$_3$/SrTiO$_3$ (001) heterostructures grown at a substrate temperature of (G) 923 K, (H) 973 K, (I) 1023 K. The crystallinities of (001) KTaO$_3$ thin films, as determined from the θ-2θ and rocking curve measurements, increase with increasing substrate temperatures from 923 K to 1023 K.

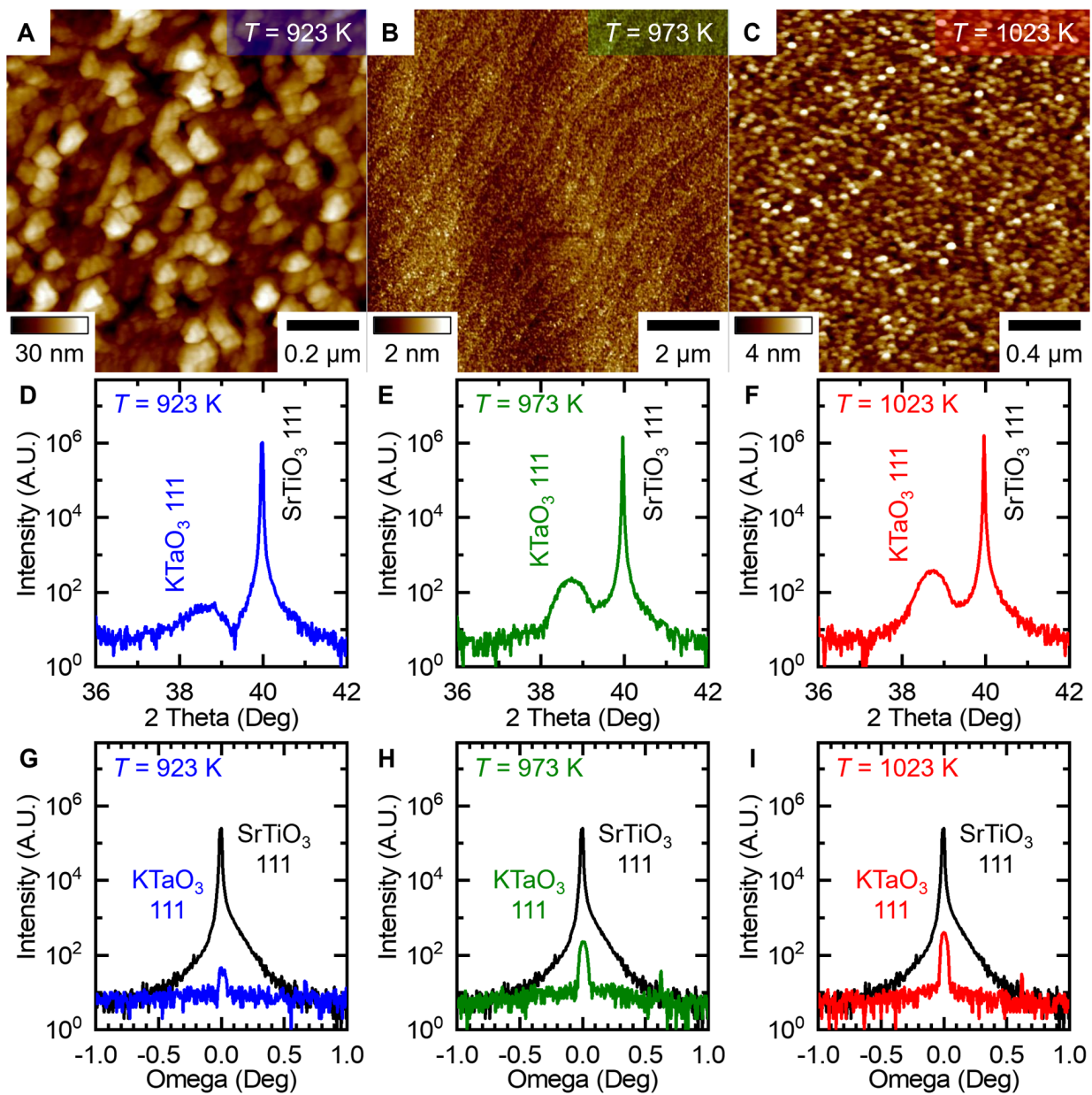

**Fig. S6. Structural characterizations of KTaO$_3$/SrTiO$_3$ (111) heterostructures.** (**A-C**) Surface topography of ≈8 nm KTaO$_3$/SrTiO$_3$ (111) heterostructures grown at a substrate temperature of (A) 923 K, (B) 973 K, (C) 1023 K. All heterostructures are grown with the same K$_2$O source temperatures of 750 K. Contrary to (001) KTaO$_3$, (111) KTaO$_3$ can be synthesized with smooth surface morphologies only within a narrow growth window near 973 K. (**D-F**) θ-2θ X-ray diffraction line scans of ≈8 nm KTaO$_3$/SrTiO$_3$ (111) heterostructures grown at a substrate temperature of (D) 923 K, (E) 973 K, (F) 1023 K. (**G-I**) Rocking curves about the KTaO$_3$ 111-diffraction conditions of 10 nm KTaO$_3$/SrTiO$_3$ (111) heterostructures grown at a substrate temperature of (G) 923 K, (H) 973 K, (I) 1023 K. Similar to (001) KTaO$_3$, the crystallinities of (111) KTaO$_3$ thin films, as determined from the θ-2θ and rocking curve measurements, increase with increasing substrate temperatures from 923 K to 1023 K.

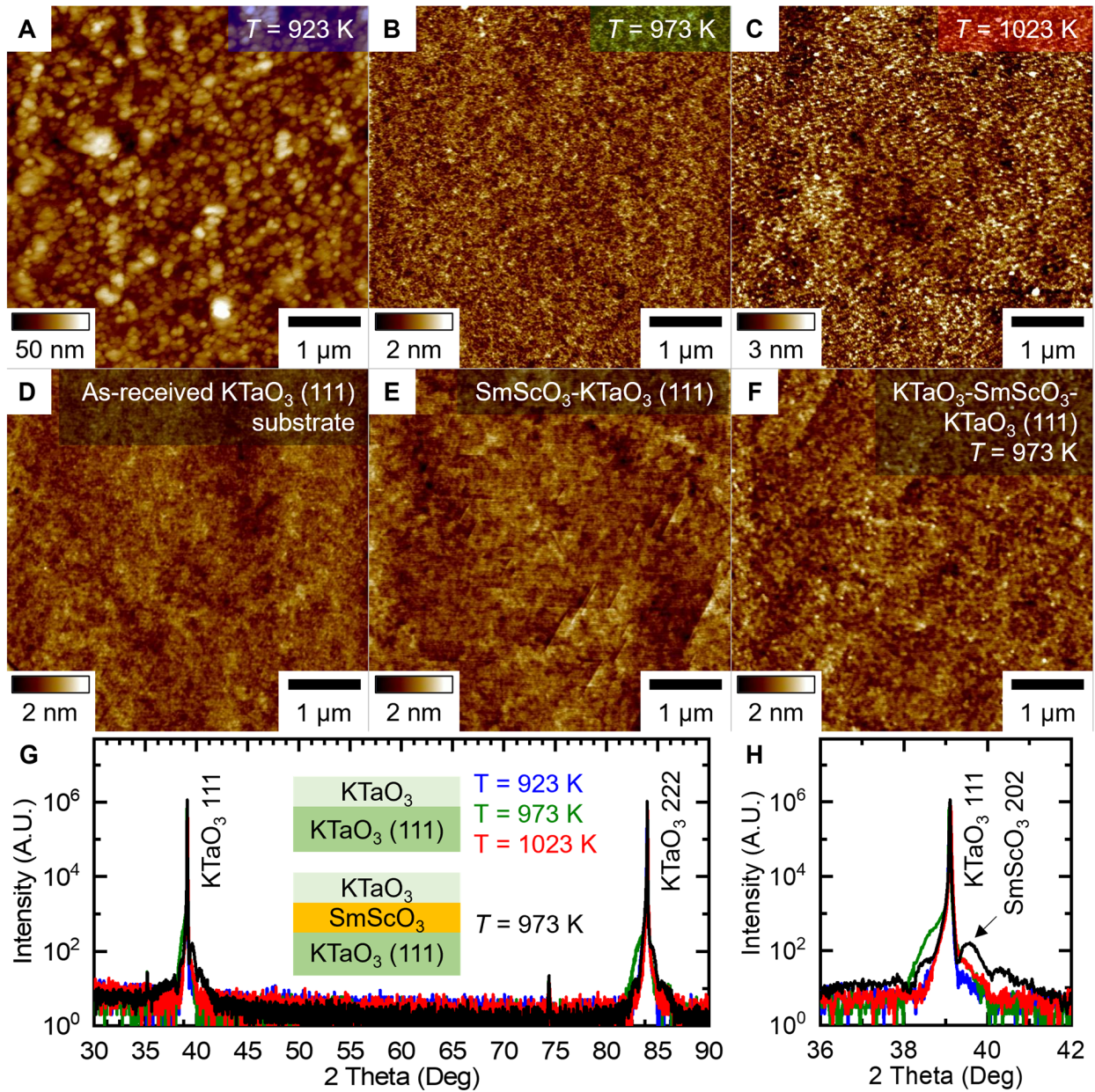

**Fig. S7. Structural characterizations of KTaO₃/KTaO₃ (111) homoepitaxial and KTaO₃/SmScO₃/KTaO₃ (111) (pseudo-)homoepitaxial structures.** (**A-C**) Surface topography of ≈8 nm KTaO₃/KTaO₃ (111) homoepitaxial structures grown at a substrate temperature of (A) 923 K, (B) 973 K, (C) 1023 K. All structures are grown with the same K₂O source temperatures of 750 K. (**D**) Surface topography of as-received KTaO₃ (111) substrate. The surface is disordered and does not show visible step-terrace structures. (**E**) Surface topography of ≈1 nm SmScO₃/KTaO₃ (111) heterostructures grown by conventional pulsed-laser deposition. (**F**) Surface topography of ≈8 nm KTaO₃/SmScO₃/KTaO₃ (111) structures. (**G**) A wide-angle θ-2θ X-ray diffraction line scans of ≈8 nm KTaO₃/KTaO₃ (111) structures grown at substrate temperatures of 923 K, 973 K, 1023 K and ≈8 nm KTaO₃/ ≈1 nm SmScO₃/KTaO₃ (111) structures grown at a substrate temperature of 973 K. (**H**) Zoom-in of (G) near the KTaO₃ 111-diffraction condition.

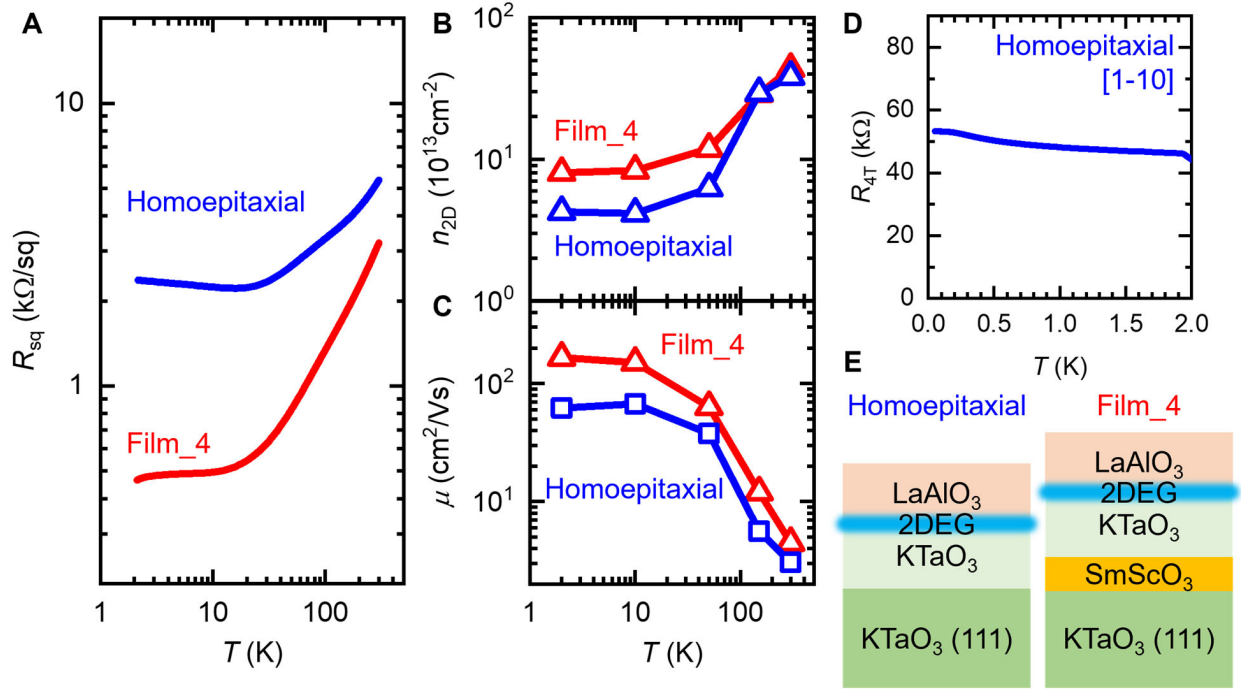

**Fig. S8. Comparison of electrical transport properties of LaAlO$_3$/KTaO$_3$/KTaO$_3$ (111) and LaAlO$_3$/KTaO$_3$/SmScO$_3$/KTaO$_3$ (111).** Temperature dependence of (**A**) $R_{sq}$, (**B**) $n_{2D}$, (**C**) $\mu$. (**D**) Temperature dependence of 4-terminal resistance measured along the [1-10] showing no superconductivity down to $T = 50$ mK. (**E**) Schematic illustrating the structures of the measured samples.

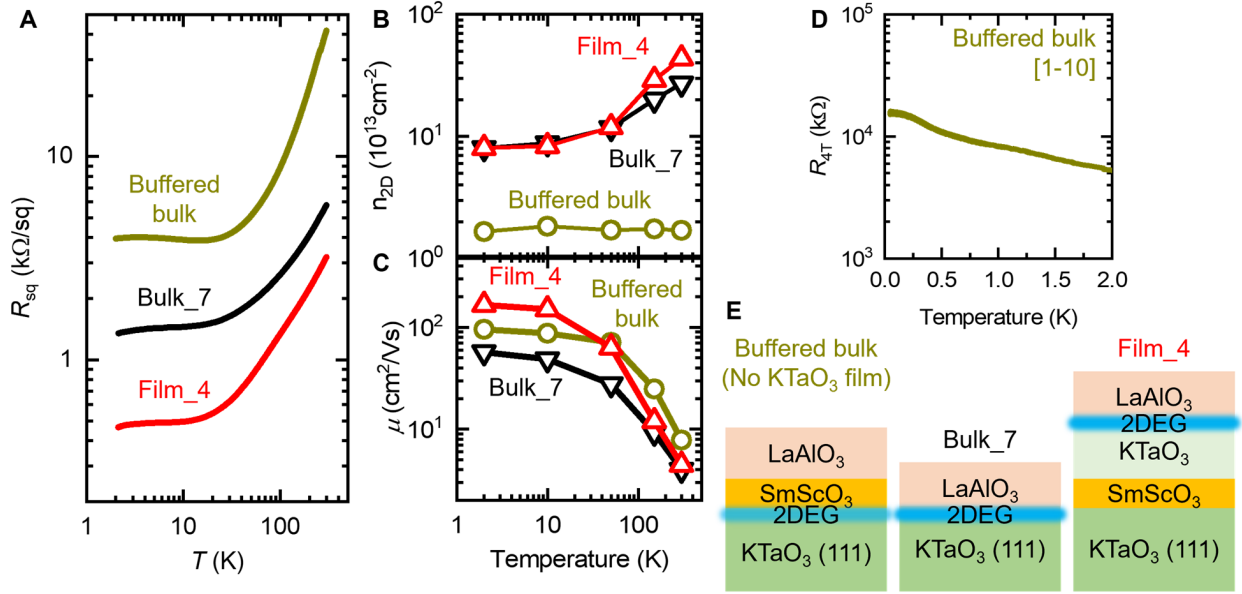

**Fig. S9. Comparison of electrical transport properties of LaAlO$_3$/SmScO$_3$/KTaO$_3$ (111), LaAlO$_3$/KTaO$_3$ (111), and LaAlO$_3$/KTaO$_3$/SmScO$_3$/KTaO$_3$ (111).** Temperature dependence of (**A**) $R_{sq}$, (**B**) $n_{2D}$, (**C**) $\mu$. (**D**) Temperature dependence of 4-terminal resistance measured along the [1-10] showing no superconductivity down to $T$ = 50 mK. (**E**) Schematic illustrating the structures of the measured samples.

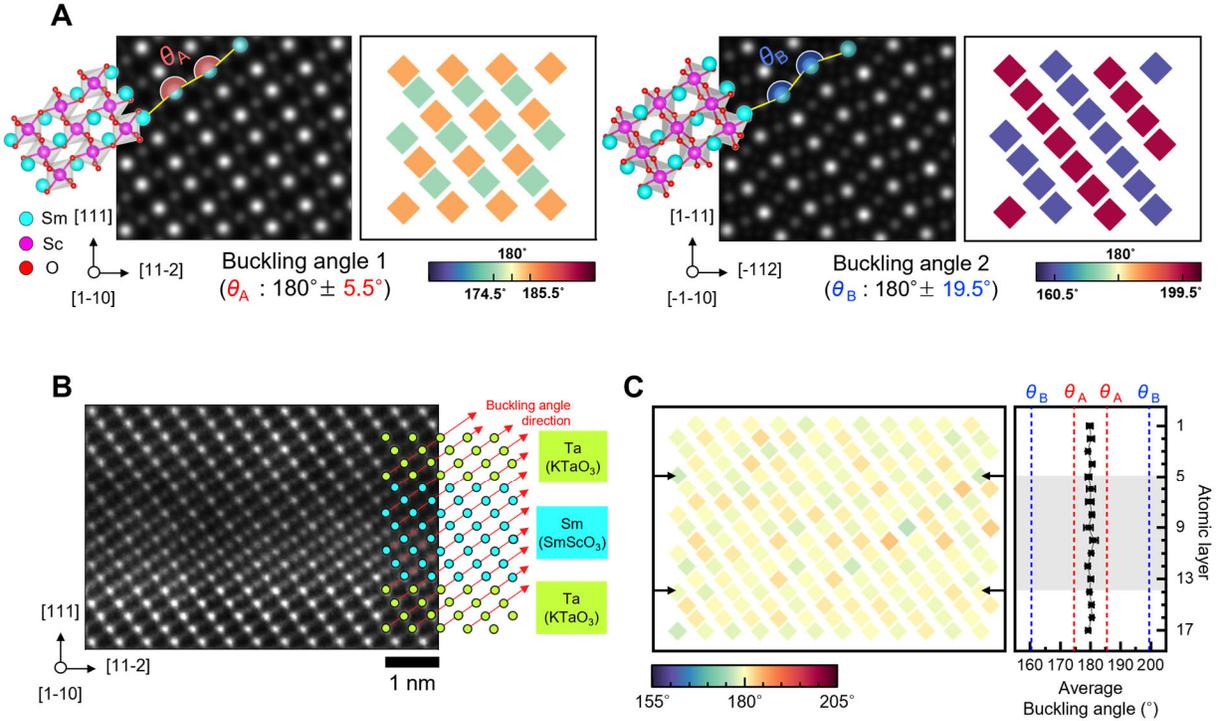

**Fig. S10. Characterization of SmScO$_3$ template layer by buckling angle.** (**A**) Simulated HAADF-STEM images and corresponding atomic models representing two types of Sm-atom buckling angle in orthorhombic SmScO$_3$ bulk state along the different crystal directions. (type 1 : $\theta_A$=180 ± 5.3°, type 2 : $\theta_B$=180 ± 19°) (**B**) Experimental HAADF-STEM image of KTaO$_3$/SmScO$_3$/KTaO$_3$ interface region. Schematic image indicates the array of atoms for buckling angle measurement along indicated direction. (Ta in KTaO$_3$, Sm in SmScO$_3$) (**C**) Buckling angle mapping and average buckling angle each atomic layer, obtained from HAADF-STEM image in (B). Comparison of two types of buckling angles in (A) ($\theta_A$ and $\theta_B$) shows that SmScO$_3$ template layer did not follow the orthorhombic bulk state.

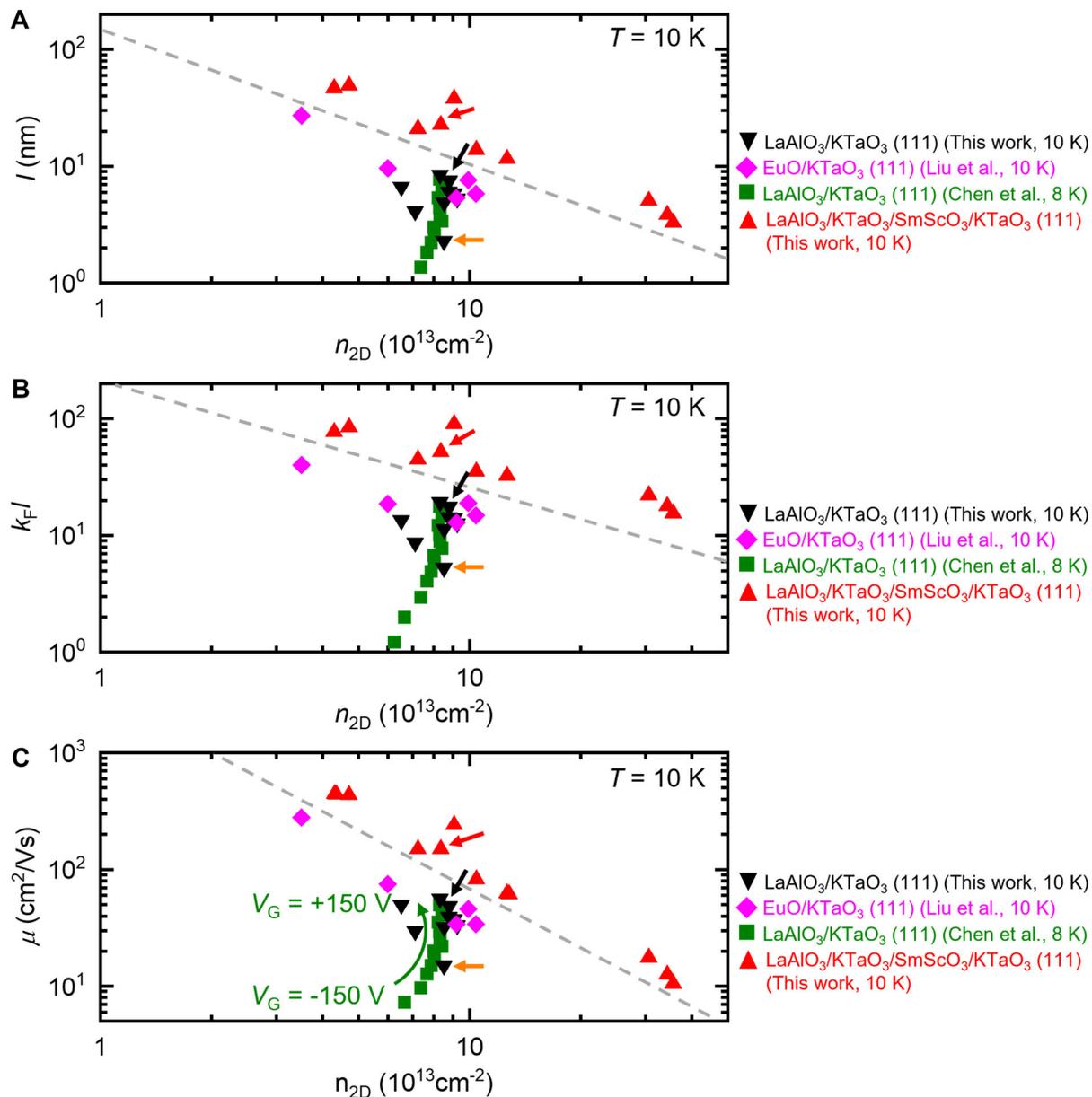

**Fig. S11. Comparison of cleanliness of various KTaO$_3$ (111) samples.** (A) $l$, (B) $k_F l$, (C) $\mu$ as a function of $n_{2D}$. Purple data are from Liu et al (*4*). Green data are back-gate voltage ($V_G$) dependence from Chen et al (*5*). All data are at $T$ = 10 K except for the green data, which are at $T$ = 8 K. Orange, black, and red arrows indicate Bulk_4, Bulk_7, and Film_4 samples.

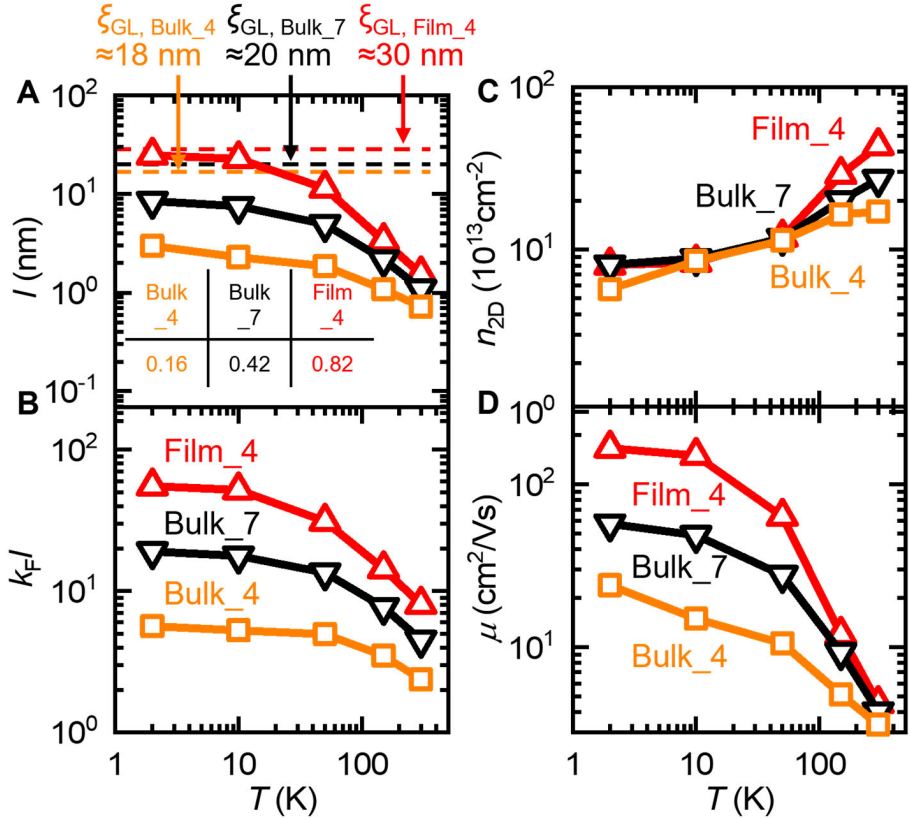

**Fig. S12. Temperature dependence of *l*, $k_Fl$, $n_{2D}$, and *μ* for Bulk_4, Bulk_7, and Film_4 samples.** Temperature dependence of (A) *l*, (B) $k_Fl$, (C) $n_{2D}$, and (D) *μ*. In (A), Ginzburg-Landau coherence length ($\xi_{GL}$) estimated from perpendicular critical field ($B_c$) is marked for reference. Inset table in (A) summarizes the $l_{2K} / \xi_{GL}$ and show that the Film_4 sample is cleaner (possessing higher $l_{2K} / \xi_{GL}$) than Bulk_4 and Bulk_7 samples.

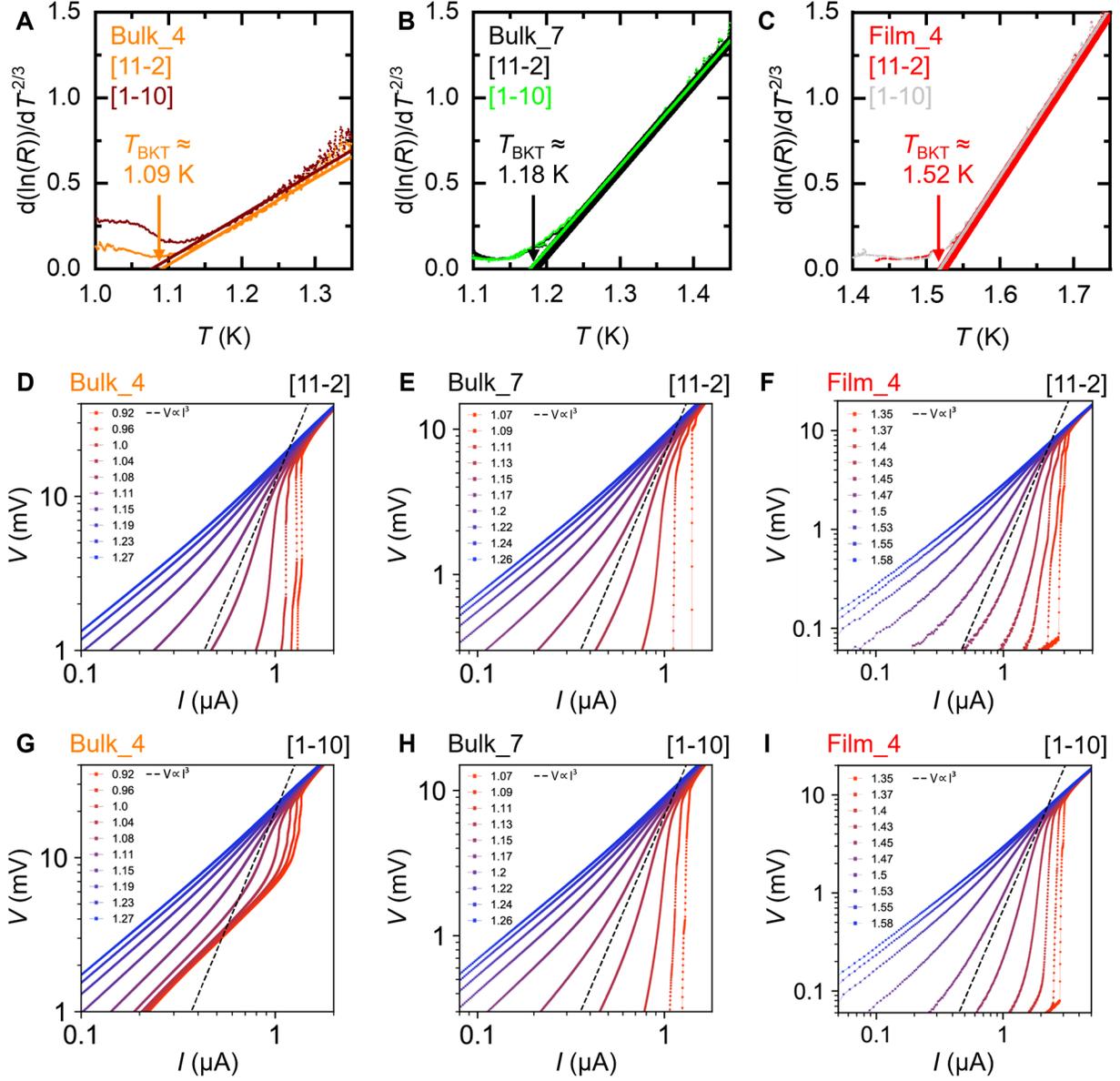

**Fig. S13. Berezinskii-Kosterlitz-Thouless (BKT) transitions.** (**A-C**) Temperature dependence of the resistance, plotted on a $[d(\ln(R))/dT]^{-2/3}$ scale, for the Bulk_4, Bulk_7 and Film_4 samples, respectively. The linear fits indicates BKT transition occurs at $T_{BKT}$ = 1.08 K, 1.18 K, and 1.52 K for the Bulk_4, Bulk_7, and Film_4 samples, respectively. (**D-I**) $V$-$J$ curves of the Bulk_4, Bulk_7 and Film_4 samples plotted on a log scale, along the [11-2] and [1-10]. Black dashed lines indicate $V \propto I^3$, which are used to infer $T_{BKT}$ = 1.09 K, 1.14 K, and 1.46 K for the Bulk_4, Bulk_7, and Film_4 samples, respectively, consistent with the results from (A-C).

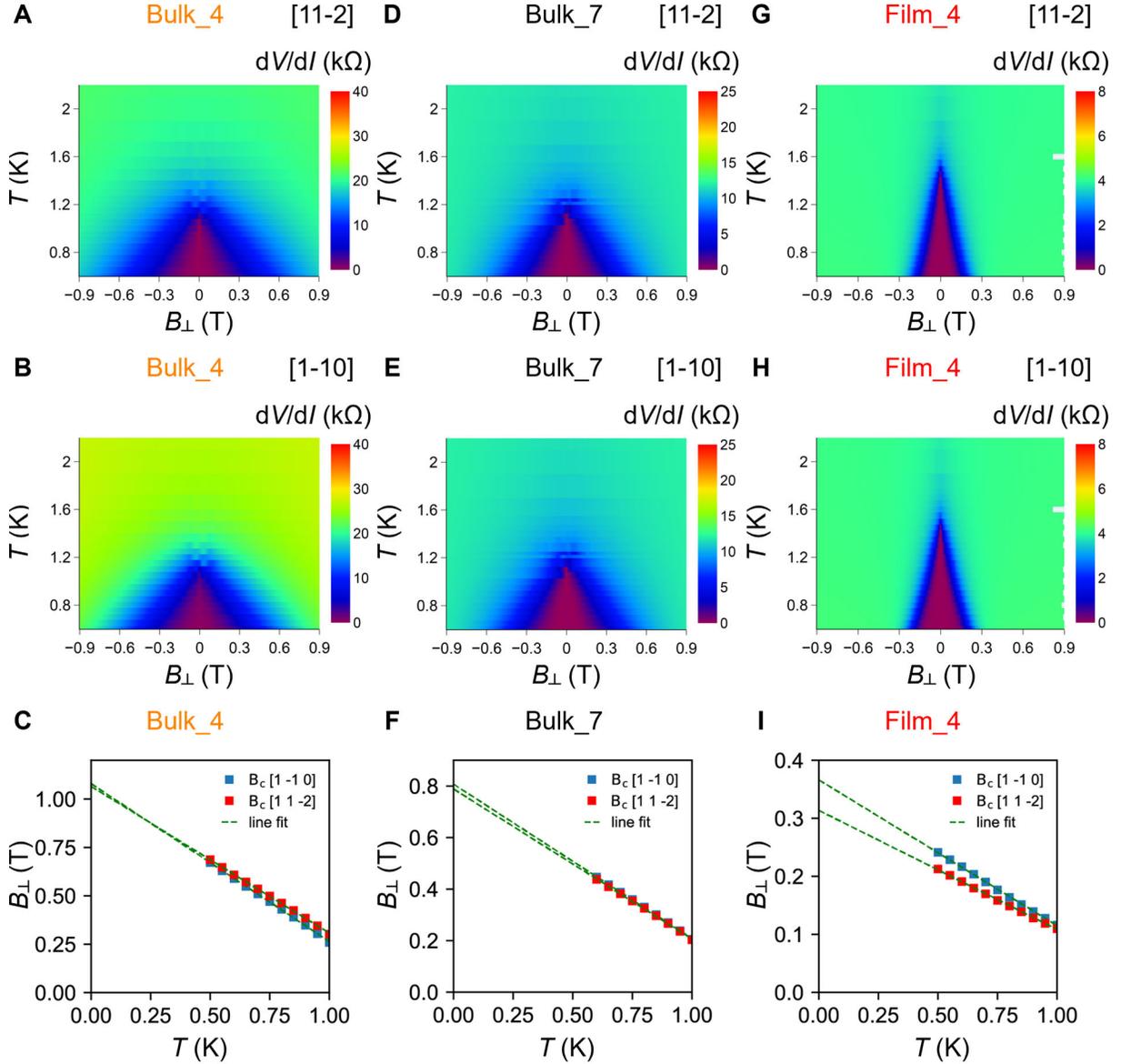

**Fig. S14. Differential resistance ($dV/dI$) as a function of $T$ and $B\perp$.** (**A** and **B**) $dV/dI$ plotted as a function of $T$ and $B\perp$ for the Bulk_4 sample along the [11-2] and [1-10], respectively. (**C**) Temperature dependence of critical field ($B_c$) of the Bulk_4 sample, defined at half of the normal state resistance ($R_N$). The linear fit indicates that at 0 K, $B_c$ along the [11-2] and [1-10] are 1.06 T and 1.08 T, giving $\xi_{GL} \approx 18$ nm. (**D** and **E**) $dV/dI$ plotted as a function of $T$ and $B\perp$ for the Bulk_7 sample along the [11-2] and [1-10], respectively. (**F**) Temperature dependence of $B_c$ of the Bulk_7 sample. The linear fit indicates that at 0 K, $B_c$ along the [11-2] and [1-10] are 0.79 T and 0.81 T, giving $\xi_{GL} \approx 20$ nm. (**G** and **H**) $dV/dI$ plotted as a function of $T$ and $B\perp$ for the Film_7 sample along the [11-2] and [1-10], respectively. (**I**) Temperature dependence of $B_c$ of the Film_7 sample. The linear fit indicates that at 0 K, $B_c$ along the [11-2] and [1-10] are 0.31 T and 0.36 T, giving $\xi_{GL} \approx 30$ nm.

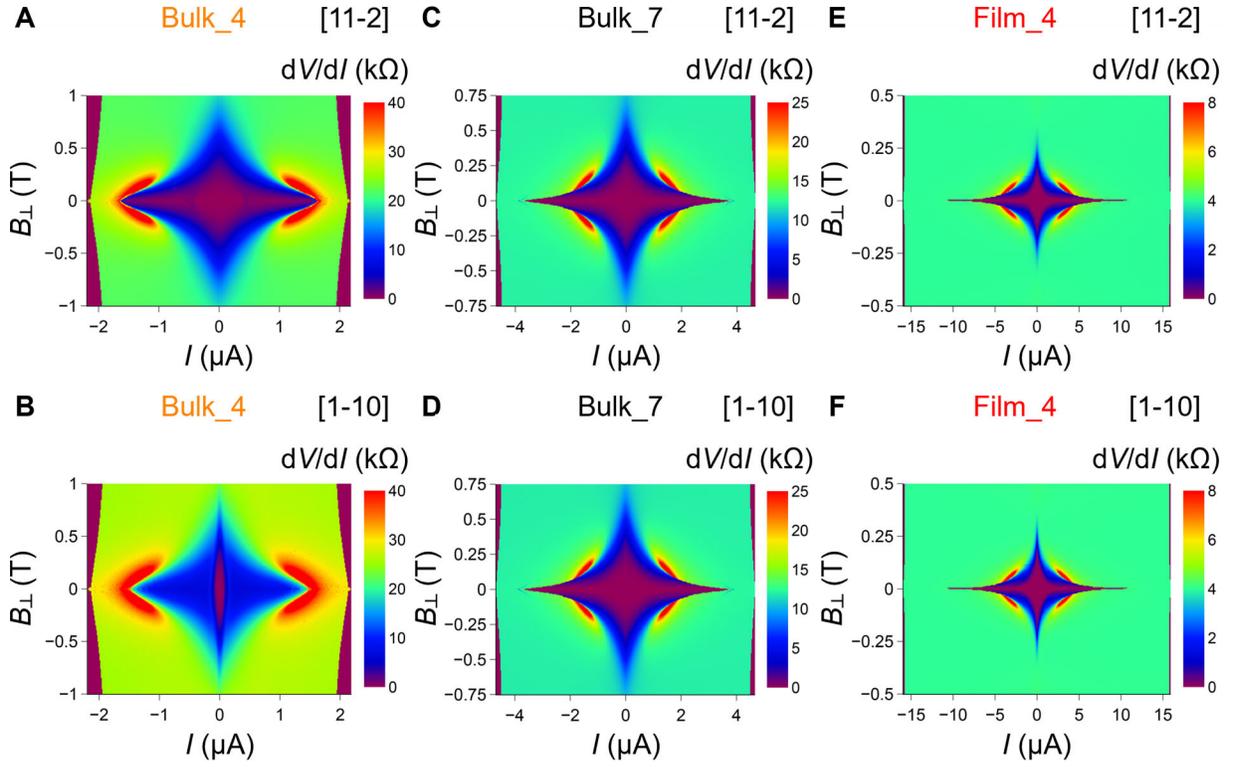

**Fig. S15. Differential resistance ($dV/dI$) as a function of $B_\perp$ and $J$.** (**A** and **B**) $dV/dI$ of the Bulk_4 sample along the [11-2] and [1-10], respectively. (**C** and **D**) $dV/dI$ of the Bulk_7 sample along the [11-2] and [1-10]. (**E** and **F**) $dV/dI$ of the Film_4 sample along the [11-2] and [1-10].

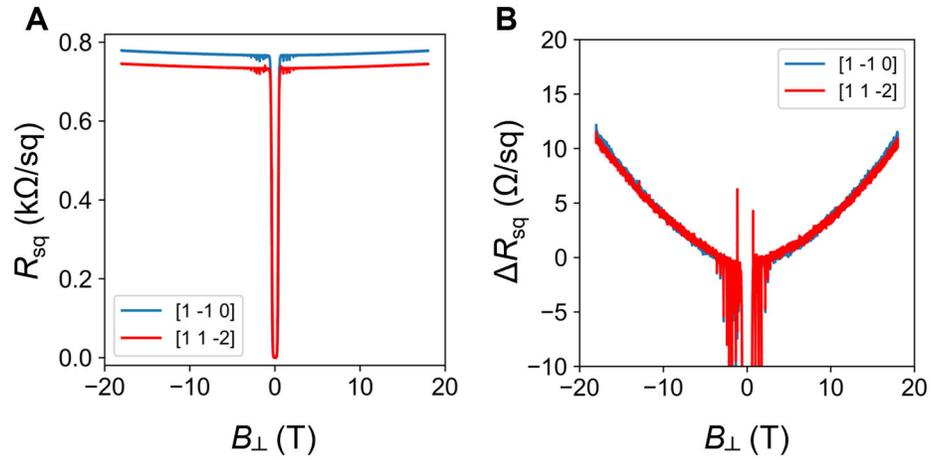

**Fig. S16. High-field magnetoresistance measurement of LaAlO$_3$/KTaO$_3$/SmScO$_3$/KTaO$_3$ heterostructures.** (**A**) Magnetic field dependence of $R_{sq}$ (along the [1-10] and [11-2]) measured at $T$ = 16 mK. (**B**) Corresponding magnetoresistance along the [1-10] and [11-2].

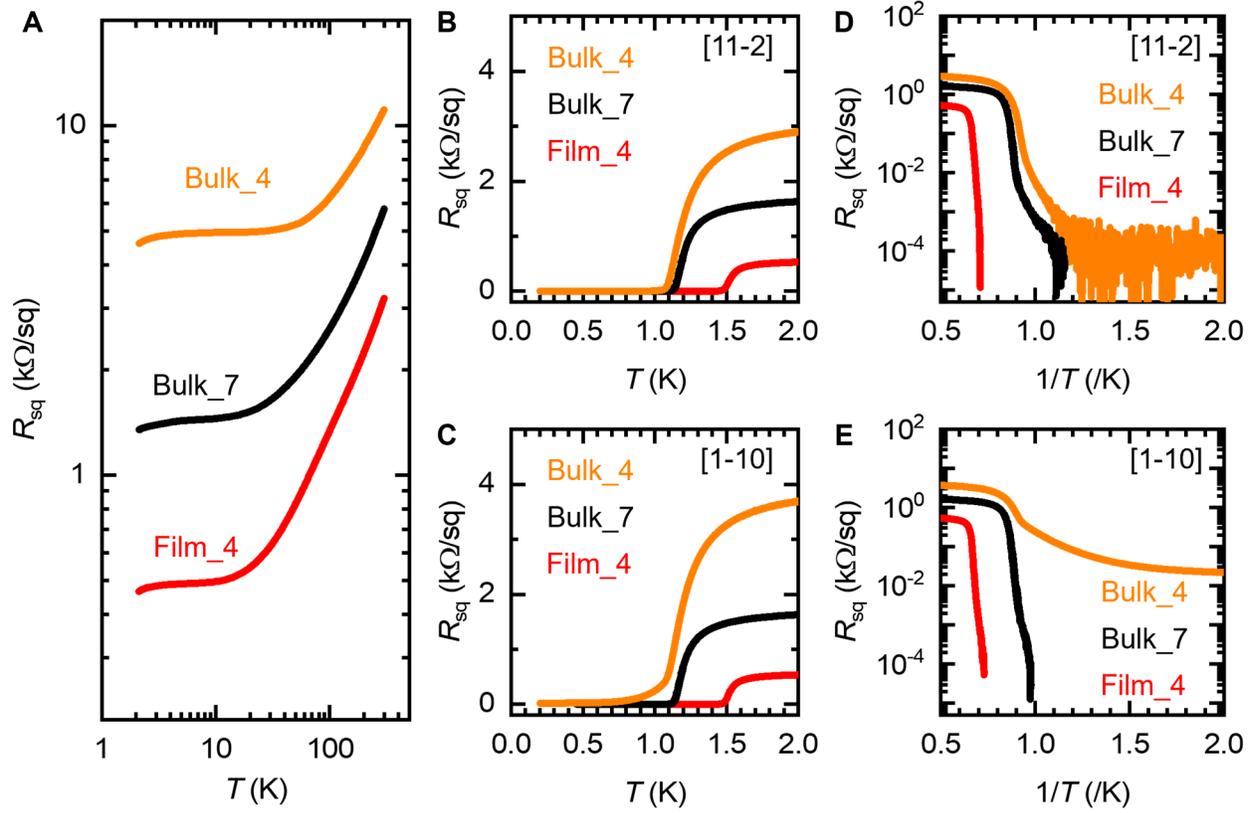

**Fig. S17. Arrhenius plots of $R_{sq}$ vs. $T$ near the superconducting transition.** Temperature dependence of (**A**) $R_{sq}$ measured with Van der Pauw method. Temperature dependence of $R_{sq}$ measured with Hall bars along the (**B**) [11-2] and (**C**) [1-10]. (**D**) Arrhenius plot of (B). (**E**) Arrhenius plot of (C).

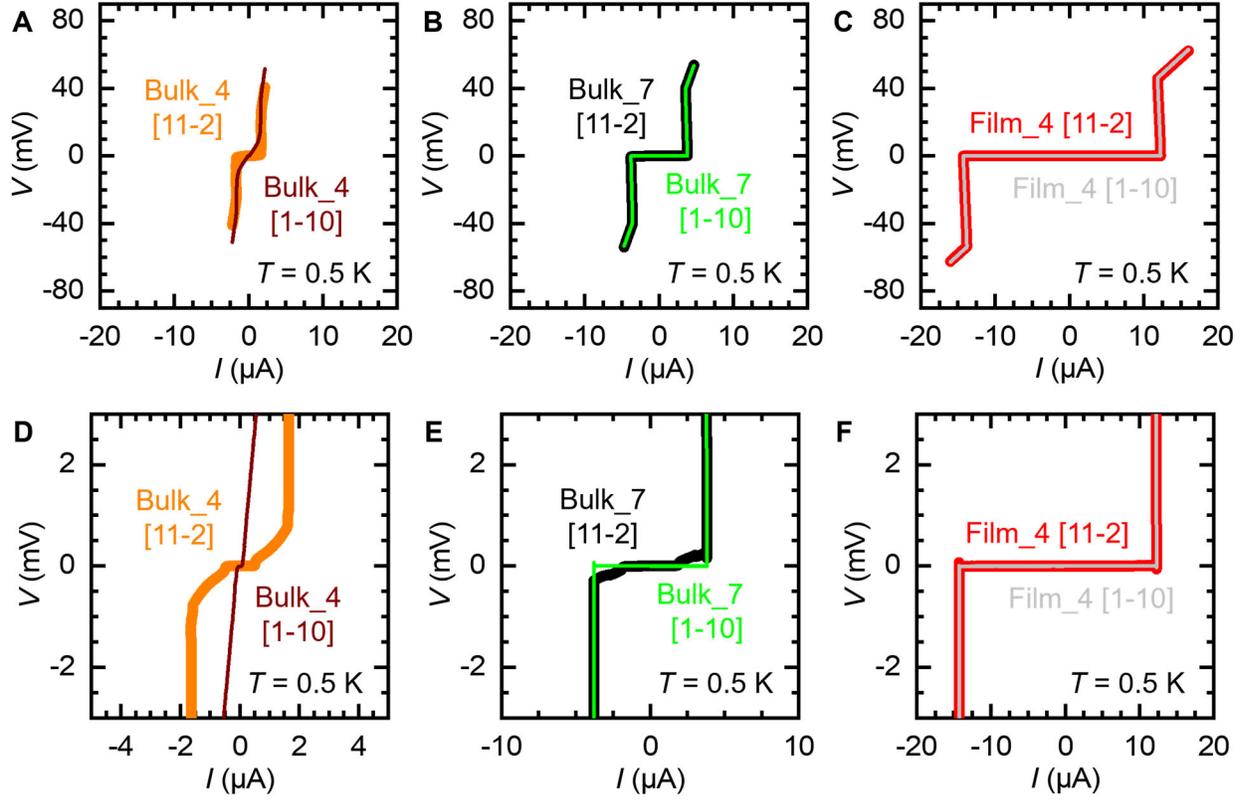

**Fig. S18.** *V-I* curves measured on Hall bars at *T* = 0.5 K. (**A**) *V-I* curves for the Bulk_4 sample along the [11-2] (orange) and [1-10] (brown). (**B**) *V-I* curves for the Bulk_7 sample along the [11-2] (black) and [1-10] (light green). (**C**) *V-I* curves for the Film_4 sample along the [11-2] (red) and [1-10] (gray). (**D-F**) Zoom-ins of (A-C) near the *V* = 0.

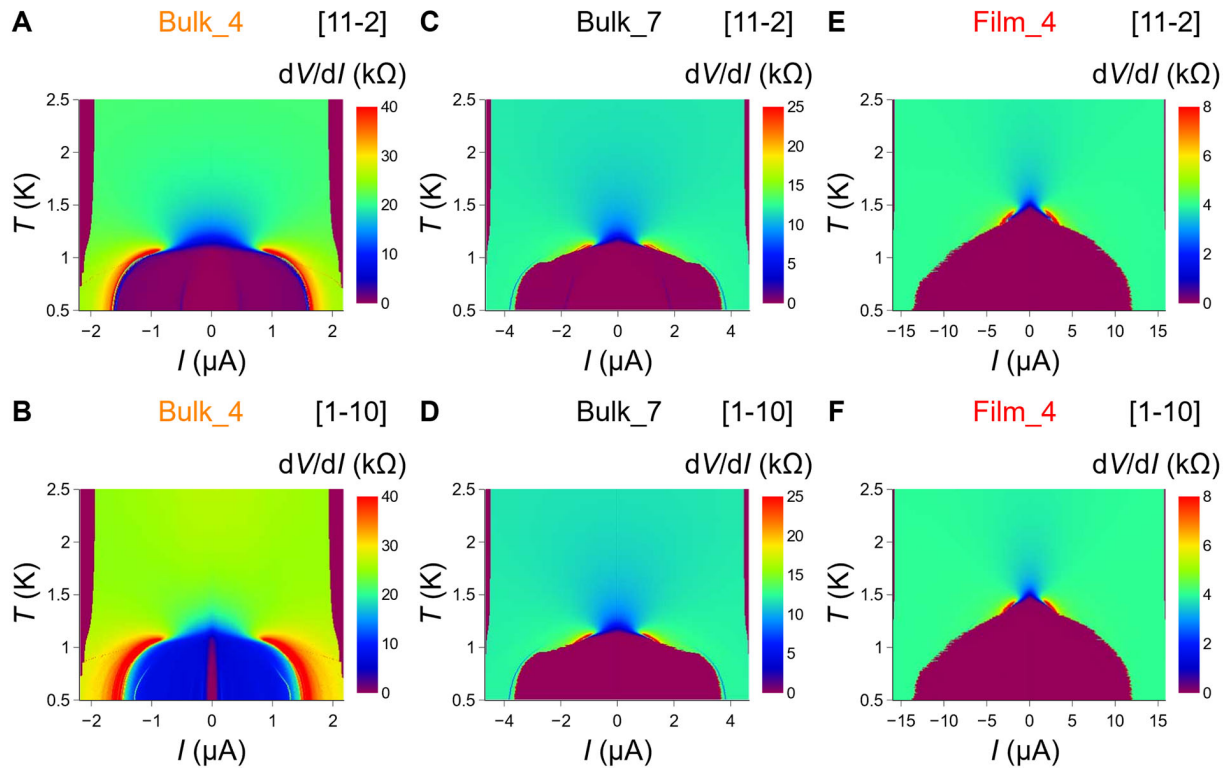

**Fig. S19. Differential resistance (d*V*/d*I*) as a function of *T* and *J*.** (**A** and **B**) d*V*/d*I* of the Bulk_4 sample along the [11-2] and [1-10], respectively. (**C** and **D**) d*V*/d*I* of the Bulk_7 sample along the [11-2] and [1-10]. (**E** and **F**) d*V*/d*I* of the Film_4 sample along the [11-2] and [1-10].

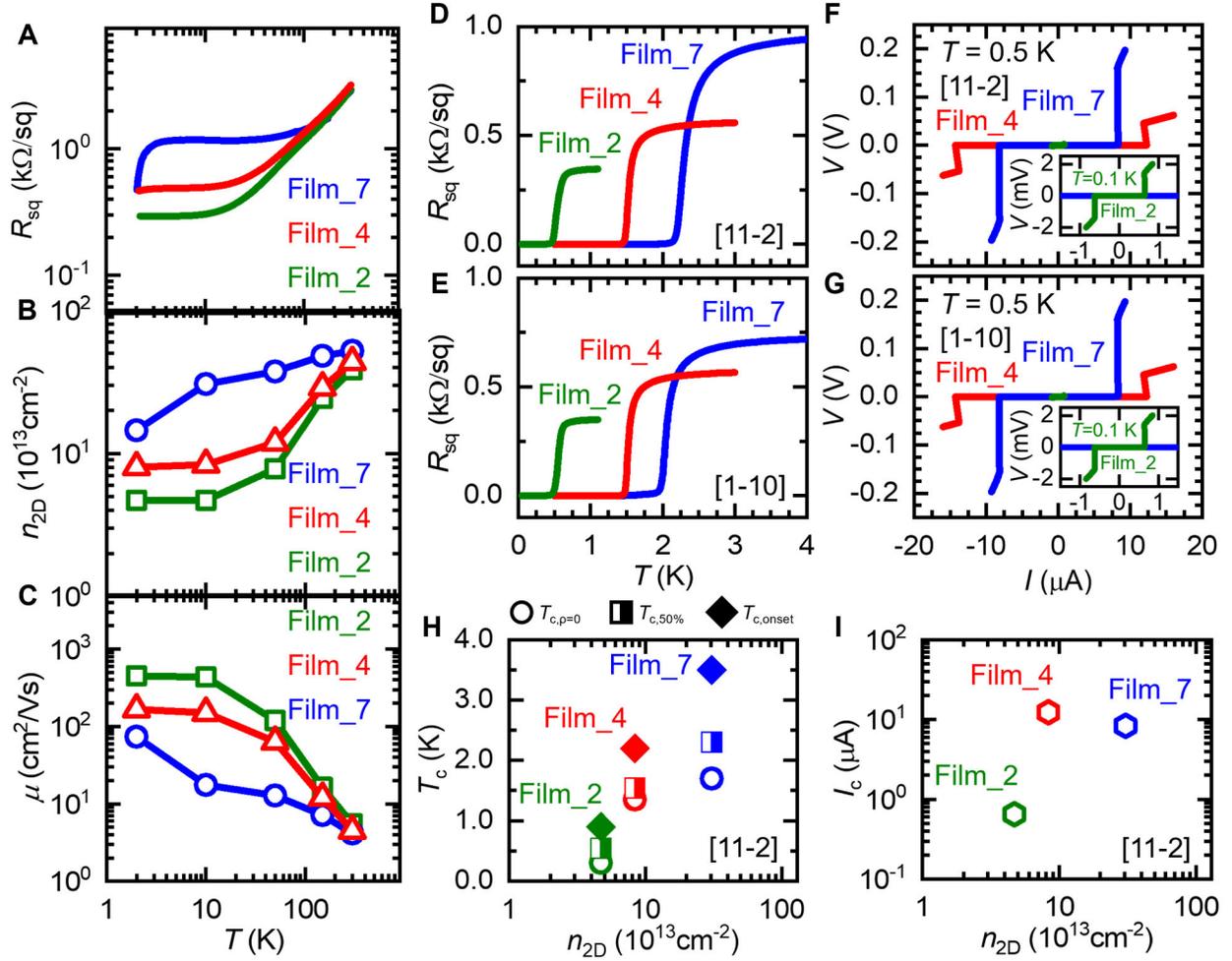

**Fig. S20. Transport properties of LaAlO$_3$/KTaO$_3$/SmScO$_3$/KTaO$_3$ (111) heterostructures with different $n_{2D}$.** Temperature dependence of (**A**) $n_{2D}$, (**B**) $\mu$, (**C**) $R_{sq}$ of the Film_2 (green), Film_4 (red), and Film_7 (blue) samples measured in a Van der Pauw geometry. Temperature dependence of $R_{sq}$ along (**D**) the [11-2], (**E**) the [1-10] measured on Hall bars. *V-I* curves along, (**F**) the [11-2], (**G**) the [1-10], measured on Hall bars at $T$ = 0.5 K (Film_4, Film_7) or $T$ = 0.1 K (Film_2). Inset shows zoom-ins near $V$ = 0. (**H**) $T_{c,\rho=0}$, $T_{c,50\%}$, and $T_{c,onset}$ vs. $n_{2D}$ based on Film_2, Film_4, and Film_7 samples. (**I**) $I_c$ vs. $n_{2D}$ based on Film_2, Film_4, and Film_7 samples. Film_2, Film_4, and Film_7 samples correspond to $n_{2D} \approx$ 8.3, 4.7, and 30.5x10$^{13}$cm$^{-2}$ at 10 K, respectively.

**Table S1. Reaction enthalpies ΔH for six ternary oxides based on the DFT-based databases of Materials Project (MP)(*48*) and the Open Quantum Materials Database (OQMD) (*49*).**

| No. | Reactions | ΔH (Joule per unit formula, J/U.F.) |
|---|---|---|
| 1 | $K_2O + 5\ Ta_2O_5 = 2\ KTa_5O_{13}$ | -219107 (for 19 atoms, by MP) |
| 2 | $K_2O + 2\ Ta_2O_5 = K_2Ta_4O_{11}$ | -383912 (for 17 atoms, by MP) |
| 3 | $K_2O + Ta_2O_5 = 2\ KTaO_3$ | -175891 (for 5 atoms, by MP)[a]<br>-179172 (for 5 atoms, by OQMD) |
| 4 | $3\ K_2O + Ta_2O_5 = 2\ K_3TaO_4$ | -336056 (for 8 atoms, by OQMD) |
| 5 | $15\ K_2O + 39\ Ta_2O_5 + 2\ Ta = 10\ K_3Ta_8O_{21}$ | -613499 (for 32 atoms, by MP) |
| 6 | $5\ K_2O + 31\ Ta_2O_5 + 13\ Ta \rightarrow 5\ K_2Ta_{15}O_{32}$ | -465582 (for 49 atoms, by MP) |

[a] This value used in the present work.

**Table S2.** Summary of sample transport properties at $T = 10$ K. Bulk_7 and Film_4 (in bold) are the same as "Bulk" and "Film" samples in Fig. 3 and Fig. 4 of the main text.

| Sample | $n_{2D}$ ($10^{13}$cm$^{-2}$) | $\mu$ (cm$^2$/Vs) | $R_{sq}$ (kΩ/sq) |
|---|---|---|---|
| Bulk_1 | 6.53 | 49.8 | 1.92 |
| Bulk_2 | 7.12 | 29.6 | 2.96 |
| Bulk_3 | 8.30 | 56.3 | 1.34 |
| Bulk_4 | 8.52 | 15.0 | 4.89 |
| Bulk_5 | 8.52 | 32.1 | 2.28 |
| Bulk_6 | 8.79 | 39.1 | 1.81 |
| **Bulk_7** | **8.81** | **48.7** | **1.46** |
| Bulk_8 | 9.07 | 37.3 | 1.84 |
| Bulk_9 | 9.25 | 33.2 | 2.03 |
| Film_1 | 4.36 | 438.9 | 0.32 |
| Film_2 | 4.71 | 434.9 | 0.30 |
| Film_3 | 7.25 | 149.67 | 0.57 |
| **Film_4** | **8.35** | **150.5** | **0.49** |
| Film_5 | 9.08 | 241.8 | 0.28 |
| Film_6 | 12.6 | 62.4 | 0.79 |
| Film_7 | 30.6 | 17.6 | 1.16 |
| Film_8 | 34.3 | 12.6 | 1.45 |
| Film_9 | 35.6 | 10.57 | 1.66 |